\documentclass[fleqn,10pt]{wlscirep}

\usepackage{amsthm}
\usepackage{bm}
\usepackage{siunitx}
\usepackage{float}
\allowdisplaybreaks

\AtBeginDocument{\providecommand{\hbar}{}\renewcommand{\hbar}{\hslash}}

\theoremstyle{definition}
\newtheorem{definition}{Definition}[section]

\theoremstyle{plain}

\newcommand{\SSC}{\mathcal{S}}
\newcommand{\CAV}{\mathcal{A}}
\newcommand{\EVC}{\mathcal{V}}
\newcommand{\UCQM}{\mathcal{U}}
\newcommand{\RED}{\mathrm{RED}}
\newcommand{\BOT}{\mathrm{BOT}}

\title{UCQM: A Six-Metric Quality Framework for Continuous-Variable Cluster States}

\author[1,*]{Saman Sarshar}

\affil[1]{ICQTs}

\affil[*]{Corresponding author: \href{mailto:saman.sarshar@shahroodut.ac.ir}{saman.sarshar@shahroodut.ac.ir}}

\begin{abstract}
Continuous-variable (CV) cluster states constitute one of the central resources for measurement-based quantum computation (MBQC). Despite substantial progress in their theoretical development and experimental realization, comparing the quality of different cluster-state topologies remains challenging, as existing approaches typically rely either on qualitative inspection of covariance matrices or on individual metrics that characterize only a single aspect of the underlying correlation structure. In this work, we propose a quantitative evaluation framework that integrates six complementary descriptors of CV cluster states: total correlation strength (SSC), correlation uniformity (CAV), error resilience (EVC), communication overhead (COM), path redundancy (RED), and bottleneck vulnerability (BOT). These quantities are combined into a single \textbf{Unified Cluster Quality Metric (UCQM)}, providing a consistent basis for evaluating and comparing different cluster-state architectures.

The proposed framework is applied to four-mode path, square, and star cluster topologies over the squeezing range $r \in [0.2,1.8]$. Across the investigated parameter regime, the square topology consistently achieves the highest UCQM score, indicating the most balanced structural characteristics among the topologies considered and supporting its suitability for measurement-based quantum computation. Because all six metrics are derived directly from covariance-matrix elements together with graph connectivity, the framework naturally extends to arbitrary $N$-mode cluster states, higher-dimensional lattice geometries, and experimentally reconstructed covariance matrices. In addition, the computational complexity of UCQM evaluation scales as $\mathcal{O}(N^3)$, making the approach applicable to large-scale cluster states.

Beyond providing a single numerical score, UCQM offers a unified perspective for analyzing the structural quality of CV cluster states and establishes a practical framework for their systematic comparison, optimization, and future design.
\end{abstract}
\graphicspath{{figures/}}
\begin{document}
\flushbottom
\maketitle
\thispagestyle{empty}

\section{INTRODUCTION}
\label{sec:introduction}

\subsection{Motivation and background}

Measurement-based quantum computation (MBQC) has become one of the leading paradigms for quantum information processing by separating the preparation of quantum resources from the computational process itself. In contrast to the conventional circuit model, where quantum gates are sequentially applied to an evolving quantum state, MBQC performs computation through adaptive local measurements on a highly entangled multipartite resource, commonly referred to as a \textit{cluster state} \cite{raussendorf2001one, briegel2009measurement, nielsen2006cluster}. This distinction is not merely conceptual. It offers important practical advantages, particularly in photonic implementations where deterministic entangling gates remain technologically demanding. Once the resource state has been prepared, the computation proceeds solely through appropriately chosen measurement bases and classical feed-forward operations \cite{kok2007linear, obrien2009photonic}.

Within the continuous-variable (CV) framework, cluster states are realized as Gaussian optical states whose entanglement structure is completely characterized by their covariance matrices. Each optical mode corresponds to a vertex of the associated graph, whereas graph edges represent controlled-phase (CZ) interactions established during state preparation \cite{menicucci2006universal, gu2009quantum, zhang2006continuous}. The deterministic generation of squeezed states together with mature linear-optical technologies has made CV platforms particularly attractive for scalable implementations of measurement-based quantum computation \cite{larsen2021fault, du2025complete, asavanant2019generation}. Over the past decade, experimental demonstrations have progressed from small proof-of-principle systems to cluster states containing dozens of optical modes \cite{yoshikawa2016generation, larsen2019deterministic}, while several theoretical architectures indicate realistic routes toward resource states comprising hundreds or even thousands of modes \cite{menicucci2011temporal, alexander2014one}.

\subsection{The challenge of comparative assessment}

Although substantial progress has been achieved in generating increasingly large CV cluster states, comparatively little attention has been devoted to establishing quantitative criteria for comparing different cluster-state architectures. In practice, one frequently encounters a fundamental question: given two cluster states with different topologies, squeezing parameters, or connectivity patterns, which one provides the more suitable computational resource? Addressing this question requires evaluation methods capable of characterizing not only the overall amount of entanglement but also the manner in which quantum correlations are distributed throughout the network.

Existing approaches exhibit several important limitations.

\textit{Qualitative assessment.} In many studies, cluster-state quality is still evaluated through visual inspection of covariance-matrix heatmaps \cite{ahadi2026ccr}. Although such representations provide intuitive insight, they remain inherently subjective, lack quantitative interpretation, and rapidly lose usefulness as system size increases.

\textit{Single-number metrics.} Widely used quantities such as logarithmic negativity \cite{vidal2002computable, plenio2005logarithmic} and von Neumann entropy \cite{wehrl1978entropy} quantify the overall degree of entanglement but provide only limited information regarding its spatial organization. Consequently, two cluster states possessing similar global entanglement may exhibit substantially different structural properties and computational performance.

\textit{Limited scalability.} Several existing metrics are formulated for specific graph families or small systems and therefore do not naturally extend to arbitrary cluster-state topologies. This limitation reduces their usefulness for analyzing large-scale architectures envisioned for practical quantum computation.

\textit{Experimental accessibility.} A number of theoretical characterization methods rely on full quantum-state tomography \cite{lvovsky2009homodyne}. While highly informative, such procedures become increasingly demanding as the number of optical modes grows. Even with modern compressive-sensing techniques, the required experimental resources typically scale as \(\mathcal{O}(N^2)\), making routine characterization of large CV systems difficult.

\subsection{Our contributions}

To overcome these limitations, we develop a unified quantitative framework for evaluating continuous-variable cluster states that combines several complementary structural descriptors within a common methodology. Rather than relying on a single indicator, the proposed framework captures multiple characteristics that jointly determine the quality of a cluster state for measurement-based quantum computation. The principal contributions of this work are summarized below.

\begin{enumerate}[label=(\arabic*)]

\item We introduce six complementary metrics—SSC, CAV, EVC, COM, RED, and BOT—each designed to quantify a distinct structural or operational aspect of cluster-state quality. All six quantities are formulated directly from covariance-matrix elements together with graph adjacency relations, allowing straightforward extension to arbitrary \(N\)-mode systems.

\item We define the \textbf{Unified Cluster Quality Metric (UCQM)}, which integrates the six normalized metrics into a single quantitative score suitable for objective comparison, benchmarking, and ranking of CV cluster-state architectures.

\item We present a detailed computational-complexity analysis demonstrating that the complete evaluation procedure scales as \(\mathcal{O}(N^3)\), making the framework computationally feasible even for cluster states containing hundreds of modes.

\item We derive analytical expressions applicable to a broad range of graph families, including complete graphs, path graphs, star graphs, regular lattices, and random graph structures.

\item We validate the proposed framework through numerical studies of representative four-mode cluster-state topologies together with scaling analyses extending to systems containing up to \(N=100\) modes, illustrating both the practicality and the physical interpretability of the proposed metrics.

\end{enumerate}

\subsection{Relation to prior work}

Existing approaches successfully characterize individual properties of continuous-variable cluster states but generally focus on only one aspect of their structure or performance. To our knowledge, no previous framework simultaneously combines quantitative evaluation, scalability to large systems, multiple complementary structural descriptors, and direct applicability to experimentally reconstructed covariance matrices. Table~\ref{tab:comparison_literature} summarizes the principal differences between representative existing methods and the framework proposed in this work.

\begin{table}[htbp]
\centering
\caption{Comparison of the proposed UCQM framework with representative approaches used to characterize quantum and continuous-variable resources. The comparison concerns the scope of the diagnostic information rather than the intrinsic validity of the individual methods.}
\label{tab:comparison_literature}
\renewcommand{\arraystretch}{1.15}
\begin{tabular}{p{3.2cm}cccc}
\toprule
\textbf{Approach} & \textbf{Quantitative} & \textbf{Scalable} & \textbf{Multi-aspect} & \textbf{Experimental access} \\
\midrule
Covariance-matrix inspection & No & Limited & No & Yes \\
Logarithmic negativity & Yes & Yes & No & In principle \\
Von Neumann entropy & Yes & Yes & No & In principle \\
Nullifier analysis & Yes & Yes & Partial & Yes \\
Graph-theoretic metrics & Yes & Yes & Partial & Yes \\
\textbf{UCQM (this work)} & \textbf{Yes} & \textbf{Yes} & \textbf{Yes} & \textbf{Yes} \\
\bottomrule
\end{tabular}
\end{table}


\subsection{Structure of the Paper}
\label{sec:paper_structure}

The paper is organized as follows. Section~2 introduces the theoretical preliminaries required for the analysis, including the basic formalism of continuous-variable quantum information, Gaussian states and covariance matrices, symplectic transformations, continuous-variable cluster states, and the graph-theoretic representation adopted throughout the work. The principal mathematical quantities and conventions used in the subsequent sections are summarized in Table~\ref{tab:notation}.

Section~3 develops the six complementary metrics that constitute the proposed framework. We first introduce the normalized total correlation strength (SSC), followed by the correlation anisotropy variance (CAV), the error vulnerability coefficient (EVC), the communication overhead (COM), the path redundancy (RED), and the bottleneck vulnerability (BOT). For each metric, its mathematical definition, physical interpretation, normalization procedure, and preferred direction of optimization are discussed.

Section~4 combines the six normalized diagnostics into the Unified Cluster Quality Metric (UCQM). Both additive and multiplicative constructions are considered, with the additive form adopted as the principal figure of merit for the numerical analysis. The choice of weighting coefficients and the interpretation of the resulting unified score are also discussed.

Section~5 examines the computational complexity of the proposed framework and identifies the dominant operations involved in evaluating the individual metrics and the UCQM. This analysis provides an estimate of the computational requirements as the number of modes increases.

Section~6 discusses the experimental accessibility of the proposed framework. In particular, we describe how the state-dependent quantities can be evaluated from experimentally reconstructed covariance matrices, while the graph-dependent quantities can be obtained from the corresponding mode-connectivity structure. The effects of finite sampling, calibration uncertainty, loss, and other experimental imperfections are also considered.

Section~7 presents the numerical results and validation of the framework. We first examine representative four-mode cluster-state topologies over the considered squeezing range and subsequently investigate the behavior of the metrics for larger systems, including cluster states with up to $N=100$ modes. The results illustrate the different trade-offs between correlation strength, correlation uniformity, communication efficiency, redundancy, and robustness across the considered graph topologies.

Section~8 discusses the relevance of the proposed framework to fault-tolerant continuous-variable measurement-based quantum computation. The connection between the proposed structural diagnostics and the nullifier-based description of realistic CV cluster states is examined, together with possible extensions incorporating finite-squeezing effects, realistic noise channels, and logical error-correction schemes.

Section~9 discusses the broader implications of the proposed framework and its relationship to conventional quantum-resource and entanglement measures. Particular attention is given to the complementary information provided by graph-theoretic and covariance-based diagnostics and to the limitations of characterizing the computational usefulness of a cluster state through a single physical quantity.

Finally, Section~10 summarizes the main results of the work, discusses the principal limitations of the present formulation, and outlines possible directions for extending the framework to non-Gaussian resources, mode-dependent squeezing, realistic experimental noise, and larger-scale fault-tolerant architectures.

\section{PRELIMINARIES}
\label{sec:preliminaries}

\subsection{Continuous-Variable Quantum Information}

Continuous-variable (CV) quantum information provides an alternative framework to qubit-based quantum computation by encoding quantum information in the continuous quadrature variables of bosonic modes \cite{weedbrook2012gaussian, braunstein2005quantum}. Because these variables are naturally associated with the electromagnetic field, CV systems constitute one of the most mature experimental platforms for quantum optics and quantum information processing.

For a single bosonic mode, the canonical quadrature operators $\hat{x}$ and $\hat{p}$ satisfy the commutation relation

\begin{equation}
[\hat{x},\hat{p}]=i,
\label{eq:display_1}
\end{equation}

where $\hbar=1$ throughout this work. These operators are related to the annihilation and creation operators through

\begin{equation}
\hat{x}=\frac{1}{\sqrt2}
(\hat a+\hat a^\dagger),
\qquad
\hat p=
\frac{1}{i\sqrt2}
(\hat a-\hat a^\dagger).
\label{eq:display_2}
\end{equation}

For an $N$-mode system it is convenient to collect all quadrature operators into the vector

\begin{equation}
\hat{\bm{\zeta}}
=
(\hat{x}_1,\hat{p}_1,
\hat{x}_2,\hat{p}_2,
\ldots,
\hat{x}_N,\hat{p}_N)^T
\in\mathbb{R}^{2N}.
\label{eq:display_3}
\end{equation}

The canonical commutation relations can then be written compactly as

\begin{equation}
[\hat{\zeta}_i,\hat{\zeta}_j]
=
i\Omega_{ij},
\label{eq:display_4}
\end{equation}

where the symplectic matrix is

\begin{equation}
\Omega
=
\bigoplus_{k=1}^{N}\omega,
\qquad
\omega=
\begin{pmatrix}
0&1\\
-1&0
\end{pmatrix}.
\label{eq:display_5}
\end{equation}

This compact notation considerably simplifies the description of multimode Gaussian states and forms the mathematical basis for the covariance-matrix formalism adopted throughout this work.


\subsection{Gaussian States and Their Characterization}

The CV cluster states considered in this paper belong to the class of Gaussian states. A quantum state is Gaussian whenever its Wigner function has a Gaussian form, or equivalently, when it is completely specified by its first and second statistical moments \cite{weedbrook2012gaussian,eisert2003introduction}. Owing to this property, the covariance matrix provides a complete description of the state and greatly simplifies both theoretical analysis and experimental characterization.

The first moments are collected in the displacement vector

\begin{equation}
\mathbf d
=
\langle
\hat{\bm{\zeta}}
\rangle,
\label{eq:display_6}
\end{equation}

whereas the second moments are contained in the covariance matrix

\begin{equation}
V_{ij}
=
\frac12
\left<
\{
\hat\zeta_i,
\hat\zeta_j
\}
\right>
-
\langle
\hat\zeta_i
\rangle
\langle
\hat\zeta_j
\rangle,
\label{eq:display_7}
\end{equation}

with $\{\cdot,\cdot\}$ denoting the anticommutator.

Throughout this work we restrict our attention to zero-displacement Gaussian states, including squeezed-vacuum and CV cluster states. Under this assumption, the covariance matrix alone completely determines all physical properties relevant to our analysis.

For every physical Gaussian state, the covariance matrix satisfies three fundamental conditions:

\begin{itemize}

\item
\textbf{Symmetry.}
By construction,

\begin{equation}
V=V^{T}.
\label{eq:display_8}
\end{equation}

\item
\textbf{Positive definiteness.}
Physical covariance matrices satisfy

\begin{equation}
V>0.
\label{eq:display_9}
\end{equation}

\item
\textbf{Robertson--Schr\"odinger uncertainty relation.}
The covariance matrix must fulfill

\begin{equation}
V+\frac{i}{2}\Omega\ge0,
\label{eq:display_10}
\end{equation}

or equivalently,

\begin{equation}
V\ge
\frac{i}{2}\Omega,
\label{eq:display_11}
\end{equation}

where the inequality is understood in the sense of positive semidefinite complex matrices.

\end{itemize}

These conditions guarantee that the covariance matrix represents a physically realizable multimode Gaussian quantum state and provide the constraints that underlie all subsequent metric definitions.

\subsection{Continuous-Variable Cluster States}

Continuous-variable cluster states constitute the fundamental resource for measurement-based quantum computation in the Gaussian regime. Mathematically, a CV cluster state is represented by a weighted graph
\begin{equation}
G=(V,E),
\label{eq:display_12}
\end{equation}
whose vertices correspond to optical modes and whose edges specify the controlled-phase (CZ) interactions responsible for generating multipartite entanglement.

More explicitly,

\begin{itemize}
    \item $V=\{1,2,\ldots,N\}$ denotes the set of vertices representing the $N$ optical modes.

    \item $E\subseteq V\times V$ denotes the set of undirected edges defining the pairwise CZ interactions between connected modes.
\end{itemize}

The graph completely determines the connectivity of the cluster state and therefore plays a central role in both its physical properties and its computational capabilities. Once the graph topology has been specified, the corresponding covariance matrix follows directly from the underlying symplectic transformation.

The preparation of a CV cluster state can be naturally divided into two successive stages \cite{menicucci2006universal,zhang2006continuous}. Independent squeezed-vacuum modes are first generated and subsequently coupled through controlled-phase interactions that imprint the desired graph connectivity onto the multimode quantum state.

\subsubsection{Step 1: Preparation of Independent Squeezed States}

Each optical mode is initially prepared in a momentum-squeezed vacuum state. For a single mode with squeezing parameter $r$, the corresponding covariance matrix is

\begin{equation}
V_{\mathrm{sq}}^{(1)}
=
\frac12
\begin{pmatrix}
e^{2r} & 0\\
0 & e^{-2r}
\end{pmatrix}.
\label{eq:display_13}
\end{equation}

Assuming that the individual modes are initially uncorrelated, the covariance matrix of the entire $N$-mode system is simply the direct sum of the single-mode covariance matrices,

\begin{equation}
V_0
=
\bigoplus_{k=1}^{N}
V_{\mathrm{sq}}^{(1)}
=
\frac12
\begin{pmatrix}
e^{2r}\mathbf I_N & 0\\
0 & e^{-2r}\mathbf I_N
\end{pmatrix},
\label{eq:display_14}
\end{equation}

where $\mathbf I_N$ denotes the $N\times N$ identity matrix.

The squeezing parameter provides a direct measure of the available quantum resource and is commonly expressed in decibels according to

\begin{equation}
r
=
\frac{\ln(10)}{20}
\times
(\mathrm{squeezing\;in\;dB})
\approx
0.1151
\times
(\mathrm{squeezing\;in\;dB}).
\label{eq:display_15}
\end{equation}

Typical experimentally accessible values range from approximately
$r=0.34$ (about $3\,\mathrm{dB}$ of squeezing) to
$r=1.84$ (approximately $16\,\mathrm{dB}$), with higher squeezing generally leading to stronger quantum correlations within the resulting cluster state \cite{vahlbruch2016ultra}.

\subsubsection{Step 2: Generation of the Cluster State via Controlled-Phase Gates}

Multipartite entanglement is established by applying controlled-phase (CZ) interactions between pairs of modes connected in the underlying graph.

For two modes $i$ and $j$, the CZ interaction transforms the quadrature operators according to

\begin{equation}
\hat{x}_i
\rightarrow
\hat{x}_i,
\qquad
\hat{p}_i
\rightarrow
\hat{p}_i+\hat{x}_j,
\label{eq:display_16}
\end{equation}

\begin{equation}
\hat{x}_j
\rightarrow
\hat{x}_j,
\qquad
\hat{p}_j
\rightarrow
\hat{p}_j+\hat{x}_i.
\label{eq:display_17}
\end{equation}

Within the symplectic formalism, the corresponding transformation is represented by

\begin{equation}
S_{\mathrm{CZ}}^{(ij)}
=
\begin{pmatrix}
\mathbf I_{N} & 0\\
A_{ij} & \mathbf I_{N}
\end{pmatrix},
\label{eq:display_18}
\end{equation}

where the matrix $A_{ij}$ contains nonzero entries only for the interacting pair of modes.

Since all CZ operations commute, the complete cluster-state preparation can be described by a single global symplectic transformation,

\begin{equation}
S
=
\prod_{(i,j)\in E}
S_{\mathrm{CZ}}^{(ij)}
=
\begin{pmatrix}
\mathbf I_N & 0\\
A & \mathbf I_N
\end{pmatrix},
\label{eq:display_19}
\end{equation}

where $A$ denotes the adjacency matrix associated with the graph $G$.

This compact representation establishes a direct correspondence between the graph topology and the Gaussian transformation that generates the cluster state. Consequently, the structural properties of the graph are reflected explicitly in the covariance matrix derived in the following subsection.

\subsection{Derivation of the Final Covariance Matrix}

Having established both the initial Gaussian resource and the symplectic transformation associated with the cluster-state graph, the covariance matrix of the final multimode state follows directly from the standard transformation rule for Gaussian states,

\begin{equation}
V=SV_{0}S^{T}.
\label{eq:display_20}
\end{equation}

To obtain an explicit expression, we first write the initial covariance matrix in block form,

\begin{equation}
V_{0}
=
\frac12
\begin{pmatrix}
X_{0} & 0\\
0 & P_{0}
\end{pmatrix},
\label{eq:display_21}
\end{equation}

where

\begin{equation}
X_{0}=e^{2r}\mathbf I_{N},
\qquad
P_{0}=e^{-2r}\mathbf I_{N}.
\label{eq:display_22}
\end{equation}

Substituting the global symplectic transformation,

\begin{equation}
S=
\begin{pmatrix}
\mathbf I_{N} & 0\\
A & \mathbf I_{N}
\end{pmatrix},
\label{eq:display_23}
\end{equation}

into the above expression gives

\begin{equation}
V
=
\frac12
\begin{pmatrix}
\mathbf I &0\\
A&\mathbf I
\end{pmatrix}
\begin{pmatrix}
X_{0}&0\\
0&P_{0}
\end{pmatrix}
\begin{pmatrix}
\mathbf I&A^{T}\\
0&\mathbf I
\end{pmatrix}.
\label{eq:display_24}
\end{equation}

Multiplication of the first two matrices yields

\begin{equation}
\begin{pmatrix}
X_{0}&0\\
AX_{0}&P_{0}
\end{pmatrix},
\label{eq:display_25}
\end{equation}

and multiplication by the remaining factor gives

\begin{equation}
V
=
\frac12
\begin{pmatrix}
X_{0}&X_{0}A^{T}\\
AX_{0}&AX_{0}A^{T}+P_{0}
\end{pmatrix}.
\label{eq:display_26}
\end{equation}

Replacing $X_{0}$ and $P_{0}$ with their explicit expressions finally leads to

\begin{equation}
V=
\frac12
\begin{pmatrix}
e^{2r}\mathbf I_{N}
&
e^{2r}A^{T}
\\
e^{2r}A
&
e^{-2r}\mathbf I_{N}
+
e^{2r}AA^{T}
\end{pmatrix},
\label{eq:cov_final}
\end{equation}

which represents the covariance matrix of an arbitrary continuous-variable cluster state generated from momentum-squeezed vacuum modes through controlled-phase interactions. Throughout the remainder of this work, Eq.~(\ref{eq:cov_final}) serves as the starting point for constructing the proposed quality metrics and analyzing their dependence on both graph topology and squeezing strength.

\subsection{Important Special Cases}

Equation~(\ref{eq:cov_final}) immediately reproduces several limiting cases that provide useful physical insight.

If the interaction graph contains no edges, the adjacency matrix satisfies

\begin{equation}
A=0,
\label{eq:display_27}
\end{equation}

and the covariance matrix reduces to

\begin{equation}
V=
\frac12
\begin{pmatrix}
e^{2r}\mathbf I_{N}&0\\
0&e^{-2r}\mathbf I_{N}
\end{pmatrix}
=
V_{0},
\label{eq:display_28}
\end{equation}

as expected for a collection of independent squeezed modes in the absence of intermode coupling.

Whenever edges are present, correlations appear in the off-diagonal blocks of the covariance matrix,

\begin{equation}
V_{x,p}
=
\frac12
e^{2r}A.
\label{eq:display_29}
\end{equation}

Consequently, for two distinct modes,

\begin{equation}
V_{x_i,p_j}
=
\frac12
e^{2r}A_{ij},
\qquad
V_{x_j,p_i}
=
\frac12
e^{2r}A_{ji}
=
\frac12
e^{2r}A_{ij},
\label{eq:display_30}
\end{equation}

where the symmetry follows directly from the undirected nature of the graph.

These off-diagonal elements quantify the position--momentum correlations established by the CZ interactions and therefore encode the entanglement structure that distinguishes one cluster-state topology from another. Since the quality metrics introduced in the following sections are derived directly from these covariance-matrix elements, Eq.~(\ref{eq:cov_final}) provides the fundamental mathematical link between graph connectivity and the quantitative characterization developed throughout this work.

\subsection{Graph Theory Background}

The quantitative framework developed in this work relies on several standard concepts from graph theory, which provide the mathematical language for describing the connectivity of continuous-variable cluster states. For completeness, we briefly summarize the definitions that will be used throughout the remainder of the paper \cite{west2001introduction,diestel2017graph}.

\begin{definition}[Adjacency Matrix]

Let
\begin{equation}
G=(V,E)
\label{eq:display_31}
\end{equation}
be an undirected graph with
\(
N=|V|
\)
vertices.

Its adjacency matrix

\begin{equation}
A\in\{0,1\}^{N\times N}
\label{eq:display_32}
\end{equation}

is defined by

\begin{equation}
A_{ij}=
\begin{cases}
1, & (i,j)\in E,\\
0, & \text{otherwise}.
\end{cases}
\label{eq:display_33}
\end{equation}

Since the graphs considered here are undirected,

\begin{equation}
A=A^{T},
\label{eq:display_34}
\end{equation}

and

\begin{equation}
A_{ii}=0,
\label{eq:display_35}
\end{equation}

indicating the absence of self-loops.

\end{definition}


\begin{definition}[Degree]

The degree of vertex $i$ is the number of edges incident to that vertex and is given by

\begin{equation}
\deg(i)
=
\sum_{j=1}^{N}
A_{ij}.
\label{eq:display_36}
\end{equation}

The corresponding degree matrix is the diagonal matrix

\begin{equation}
D
=
\mathrm{diag}
\left(
\deg(1),
\deg(2),
\ldots,
\deg(N)
\right),
\label{eq:display_37}
\end{equation}

whose diagonal entries contain the vertex degrees.

\end{definition}


\begin{definition}[Complete Graph $K_N$]

A complete graph contains every possible edge between distinct vertices,

\begin{equation}
|E|
=
\binom{N}{2},
\label{eq:display_38}
\end{equation}

so that

\begin{equation}
A_{ij}=1,
\qquad
i\neq j.
\label{eq:display_39}
\end{equation}

Every vertex therefore has degree

\begin{equation}
N-1.
\label{eq:display_40}
\end{equation}

\end{definition}


\begin{definition}[Path Graph $P_N$]

A path graph consists of the sequence of edges

\begin{equation}
(1,2),
(2,3),
\ldots,
(N-1,N).
\label{eq:display_41}
\end{equation}

Its adjacency matrix contains nonzero elements only on the first upper and lower diagonals, reflecting the nearest-neighbor connectivity of the graph.

\end{definition}


\begin{definition}[Cycle Graph $C_N$]

A cycle graph extends the path graph by connecting the two terminal vertices,

\begin{equation}
(N,1),
\label{eq:display_42}
\end{equation}

thereby forming a closed loop.

For

\begin{equation}
N=4,
\label{eq:display_43}
\end{equation}

the resulting graph corresponds to the square cluster state investigated later in this work.

\end{definition}


\begin{definition}[Star Graph $K_{1,N-1}$]

A star graph consists of one central vertex connected to all remaining vertices, while no direct connections exist among the peripheral vertices.

For

\begin{equation}
N=4,
\label{eq:display_44}
\end{equation}

this topology coincides with the T-shaped cluster state considered in our numerical analysis.

\end{definition}


\begin{definition}[Two-Dimensional Square Lattice]

A two-dimensional lattice

\begin{equation}
L_{m\times m}
\label{eq:display_45}
\end{equation}

contains

\begin{equation}
N=m^{2}
\label{eq:display_46}
\end{equation}

vertices arranged on a square grid.

Each interior vertex is connected to its nearest neighbors in the horizontal and vertical directions, while boundary vertices possess fewer neighbors owing to the finite lattice size.

Such lattice structures constitute one of the principal resource states for large-scale measurement-based quantum computation and provide a natural setting for extending the framework proposed in this work to higher-dimensional cluster-state architectures.

\end{definition}


\section{The Six Metrics}
\label{sec:metrics}

In this section, we introduce the six metrics that underpin the proposed framework for assessing continuous-variable cluster states. The central idea is that no single quantity is sufficient to characterize the usefulness of a cluster state for measurement-based quantum computation (MBQC). A state may carry a large amount of correlation but still be poorly suited for computation if those correlations are unevenly distributed, overly concentrated around a small set of nodes, or fragile under perturbations. For this reason, we consider a family of complementary descriptors, each designed to capture a distinct physical aspect of the underlying graph state.

All metrics are expressed in terms of the covariance matrix elements $V_{x_i,p_j}$, which, for the idealized cluster states considered here, are directly linked to the adjacency structure of the graph through Eq.~\eqref{eq:cov_final}. This representation provides a convenient bridge between the continuous-variable state description and the graph-theoretic properties of the cluster. In what follows, we define each metric carefully and discuss the physical intuition behind it.
Before defining the six metrics, we establish notation for the four graph families that will appear repeatedly in our numerical examples: the complete graph \(K_N\) (all possible edges), the path graph \(P_N\) (a linear chain), the star graph \(K_{1,N-1}\) (a single central hub connected to all peripheral vertices), and the two-dimensional square lattice \(L_{\sqrt{N}\times\sqrt{N}}\) (nearest-neighbor grid). These topologies represent qualitatively different regimes of connectivity and serve as benchmarks throughout the paper.

\subsection{Sum of Squared Correlations (SSC)}
\label{sec:ssc}

The Sum of Squared Correlations (SSC) is introduced as a global descriptor of the intermode correlations encoded by the graph structure of a continuous-variable cluster state. Unlike local measures that characterize individual modes or specific subgraphs, SSC quantifies the cumulative correlation strength distributed across the entire network. Since the off-diagonal position--momentum blocks of the covariance matrix are determined directly by the graph connectivity, SSC provides a quantitative measure of the overall correlation resources available within the cluster state. For a fixed squeezing parameter, larger and more densely connected graphs naturally yield larger SSC values because a greater number of mode pairs contribute to the total correlation weight.

For an $N$-mode CV cluster state with covariance matrix $V$, the SSC is defined as

\begin{equation}
\SSC(G)=
\frac12
\sum_{i=1}^{N}
\sum_{j=1}^{N}
\left(
V_{x_i,p_j}
\right)^2,
\label{eq:ssc_def}
\end{equation}

where the prefactor $1/2$ compensates for the double counting of symmetric mode pairs in undirected graphs.

Using the covariance-matrix representation derived in Eq.~\eqref{eq:cov_final},

\begin{equation}
V_{x_i,p_j}
=
\frac12
e^{2r}
A_{ij},
\label{eq:display_47}
\end{equation}

the SSC becomes

\begin{equation}
\SSC(G)
=
\frac12
\sum_{i=1}^{N}
\sum_{j=1}^{N}
\left(
\frac12
e^{2r}
A_{ij}
\right)^2
=
\frac18
e^{4r}
\sum_{i=1}^{N}
\sum_{j=1}^{N}
A_{ij}^{2}.
\label{eq:display_48}
\end{equation}

For simple graphs,

\begin{equation}
A_{ij}^{2}=A_{ij},
\label{eq:display_49}
\end{equation}

and the double summation counts every undirected edge twice,

\begin{equation}
\sum_{i,j}
A_{ij}
=
2|E|.
\label{eq:display_50}
\end{equation}

Substituting these identities yields the compact expression

\begin{equation}
\SSC(G)
=
\frac14
e^{4r}
|E|,
\label{eq:ssc_closed}
\end{equation}

demonstrating that SSC scales linearly with the number of graph edges and exponentially with the squeezing parameter.

Within the ideal covariance-matrix model of Eq.~\eqref{eq:cov_final}, SSC depends exclusively on the graph topology and the squeezing strength. It therefore provides a convenient global indicator of the correlation resources encoded in the cluster state. To compare systems of different sizes, it is useful to eliminate the trivial dependence on both $N$ and $r$ through normalization.

For a graph containing $N$ vertices, the maximum possible number of edges is achieved by the complete graph,

\begin{equation}
|E|_{\max}
=
\binom{N}{2}
=
\frac{N(N-1)}{2},
\label{eq:display_51}
\end{equation}

leading to the maximal SSC value

\begin{equation}
\SSC_{\max}(N,r)
=
\frac14
e^{4r}
\frac{N(N-1)}{2}
=
\frac{e^{4r}N(N-1)}{8}.
\label{eq:display_52}
\end{equation}

The normalized metric is therefore

\begin{equation}
\widehat{\SSC}(G)
=
\frac{\SSC(G)}
{\SSC_{\max}(N,r)}
=
\frac{2|E|}
{N(N-1)},
\label{eq:ssc_norm}
\end{equation}

which, in the ideal binary-adjacency limit, is identical to the normalized edge density of the underlying graph. The dependence on the squeezing parameter cancels exactly, allowing direct comparison between cluster states prepared under different squeezing conditions. Consequently, values of $\widehat{\SSC}$ close to unity indicate highly connected graph structures, whereas smaller values correspond to sparse topologies with weaker overall correlation support.

\begin{figure}[htbp]
\centering
\includegraphics[width=1\linewidth]{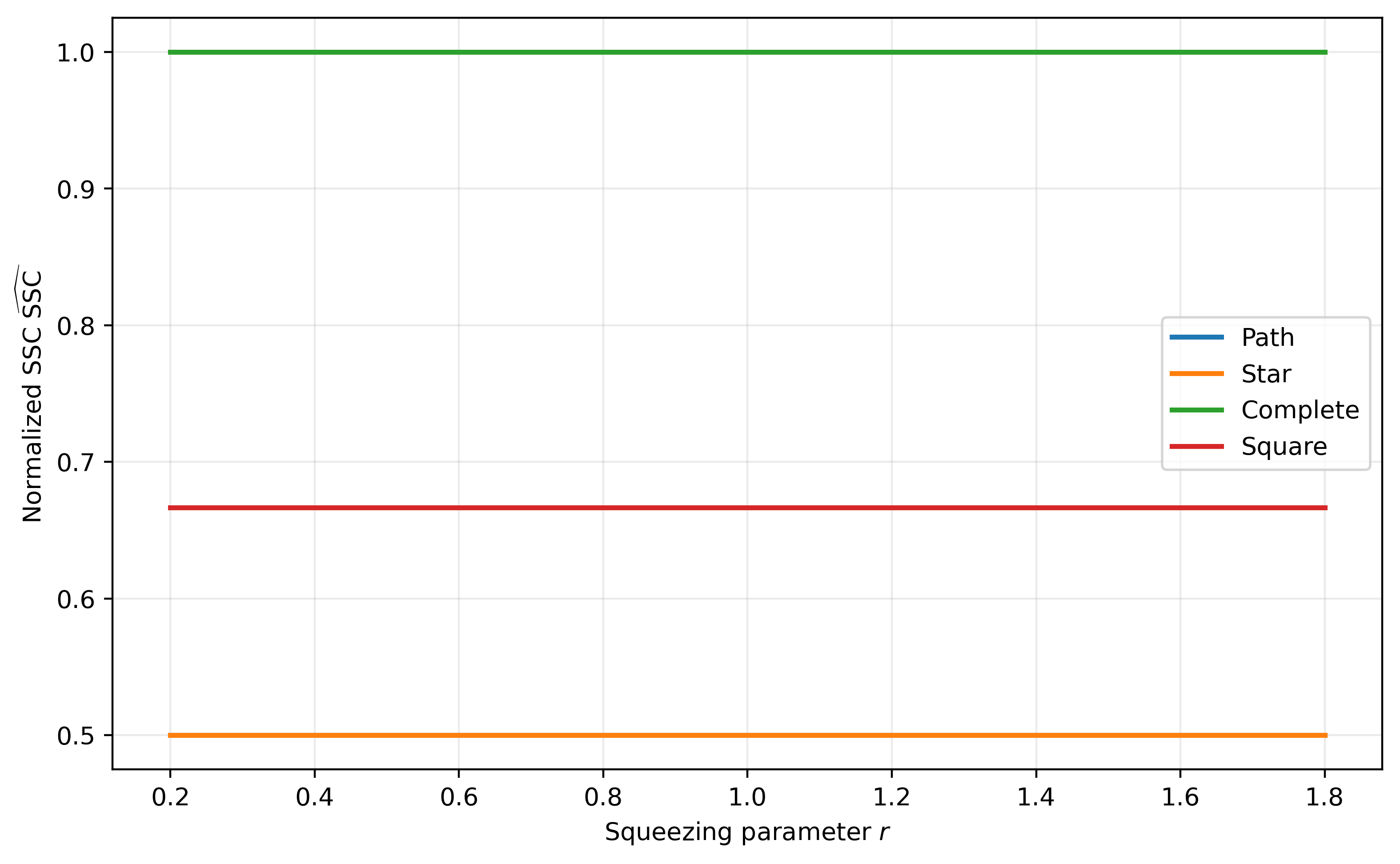}
\caption{Sum of Squared Correlations (SSC) as a function of squeezing parameter \(r\) for the topologies.}
\label{fig:ssc}
\end{figure}

\subsection{Correlation Anisotropy Variance (CAV)}
\label{sec:cav}

Whereas the Sum of Squared Correlations (SSC) quantifies the overall amount of graph-induced correlations, it provides no information about how these correlations are distributed throughout the cluster state. Two graphs may possess comparable total correlation strength while exhibiting markedly different internal organizations. In one topology, the available correlations may be distributed almost uniformly over all mode pairs, whereas in another they may be concentrated around a limited subset of highly connected regions. Such differences are particularly relevant for measurement-based quantum computation (MBQC), since an uneven distribution of correlations can lead to preferential information-flow channels and reduced resilience against localized imperfections.

To characterize this structural property, we introduce the Correlation Anisotropy Variance (CAV), which measures the degree of heterogeneity in the distribution of pairwise graph-induced correlations.

Let

\begin{equation}
M=\binom{N}{2}
\label{eq:display_53}
\end{equation}

denote the number of unordered pairs of modes. For every pair $(i,j)$, we define the effective correlation weight

\begin{equation}
W_{ij}
=
e^{2r}
\left[
A_{ij}
+
\frac12(AA^{T})_{ij}
\right],
\label{eq:display_54}
\end{equation}

where the first term represents the direct graph connectivity and the second term accounts for indirect structural correlations arising from common neighbors.

The quantity

\begin{equation}
(AA^{T})_{ij}
=
\sum_{k=1}^{N}
A_{ik}A_{jk}
\label{eq:display_55}
\end{equation}

counts the number of common neighbors shared by vertices $i$ and $j$. Consequently, unlike the adjacency matrix itself, which contains only nearest-neighbor connectivity, the product $AA^{T}$ captures second-order topological information describing the local environment surrounding each pair of vertices. This additional information enables CAV to distinguish graph topologies that possess identical edge densities but different local connectivity structures.

The second-order contribution naturally appears in the covariance matrix through the momentum block,

\begin{equation}
V_{p_i,p_j}
=
\frac12
e^{2r}
(AA^{T})_{ij},
\label{eq:display_56}
\end{equation}

derived in Eq.~\eqref{eq:cov_final}.

The Correlation Anisotropy Variance is then defined as the sample variance of the complete set of pairwise correlation weights,

\begin{equation}
\CAV(G)
=
\frac{1}{M-1}
\sum_{i<j}
\left(
W_{ij}
-
\bar W
\right)^2,
\label{eq:cav_var}
\end{equation}

where

\begin{equation}
\bar W
=
\frac1M
\sum_{i<j}
W_{ij}
\label{eq:display_57}
\end{equation}

denotes the mean pairwise correlation weight.

By construction, small values of CAV indicate that the available correlations are distributed relatively uniformly throughout the graph, whereas larger values correspond to increasingly anisotropic correlation landscapes in which the correlation resources become concentrated around specific regions or mode pairs.

Substituting the definition of $W_{ij}$ into Eq.~\eqref{eq:cav_var} yields

\begin{equation}
\CAV(G)
=
e^{4r}
\frac{1}{M-1}
\sum_{i<j}
\left(
B_{ij}
-
\bar B
\right)^2,
\label{eq:display_58}
\end{equation}

where

\begin{equation}
B_{ij}
=
A_{ij}
+
\frac12(AA^{T})_{ij},
\qquad
\bar B
=
\frac1M
\sum_{i<j}
B_{ij}.
\label{eq:display_59}
\end{equation}

This expression shows that the squeezing parameter contributes only through the overall multiplicative factor $e^{4r}$, while the purely topological information is entirely contained in the fluctuations of $B_{ij}$. Consequently, the raw CAV simultaneously reflects both the squeezing strength and the graph architecture.

To compare cluster states prepared with different squeezing levels, the trivial squeezing dependence is removed through normalization with respect to the maximal attainable value for fixed system size and squeezing,

\begin{equation}
\widehat{\CAV}(G)
=
\frac{\CAV(G)}
{\CAV_{\max}(N,r)},
\label{eq:cav_norm}
\end{equation}

where

\begin{equation}
\CAV_{\max}(N,r)
=
e^{4r}
\max_G
\left[
\frac1{M-1}
\sum_{i<j}
(B_{ij}-\bar B)^2
\right].
\label{eq:display_60}
\end{equation}

Since both the numerator and denominator contain the same global factor $e^{4r}$, the squeezing dependence cancels exactly, giving

\begin{equation}
\widehat{\CAV}(G)
=
\frac{
\displaystyle
\sum_{i<j}(B_{ij}-\bar B)^2
}{
\displaystyle
\max_G
\sum_{i<j}(B_{ij}-\bar B)^2
}.
\label{eq:display_61}
\end{equation}

The normalized metric therefore depends only on the graph topology and satisfies

\begin{equation}
0
\le
\widehat{\CAV}(G)
\le
1.
\label{eq:display_62}
\end{equation}

Values close to zero correspond to nearly isotropic correlation distributions, whereas values approaching unity indicate strongly heterogeneous correlation structures with pronounced directional imbalance.

\begin{figure}[htbp]
\centering
\includegraphics[width=\linewidth]{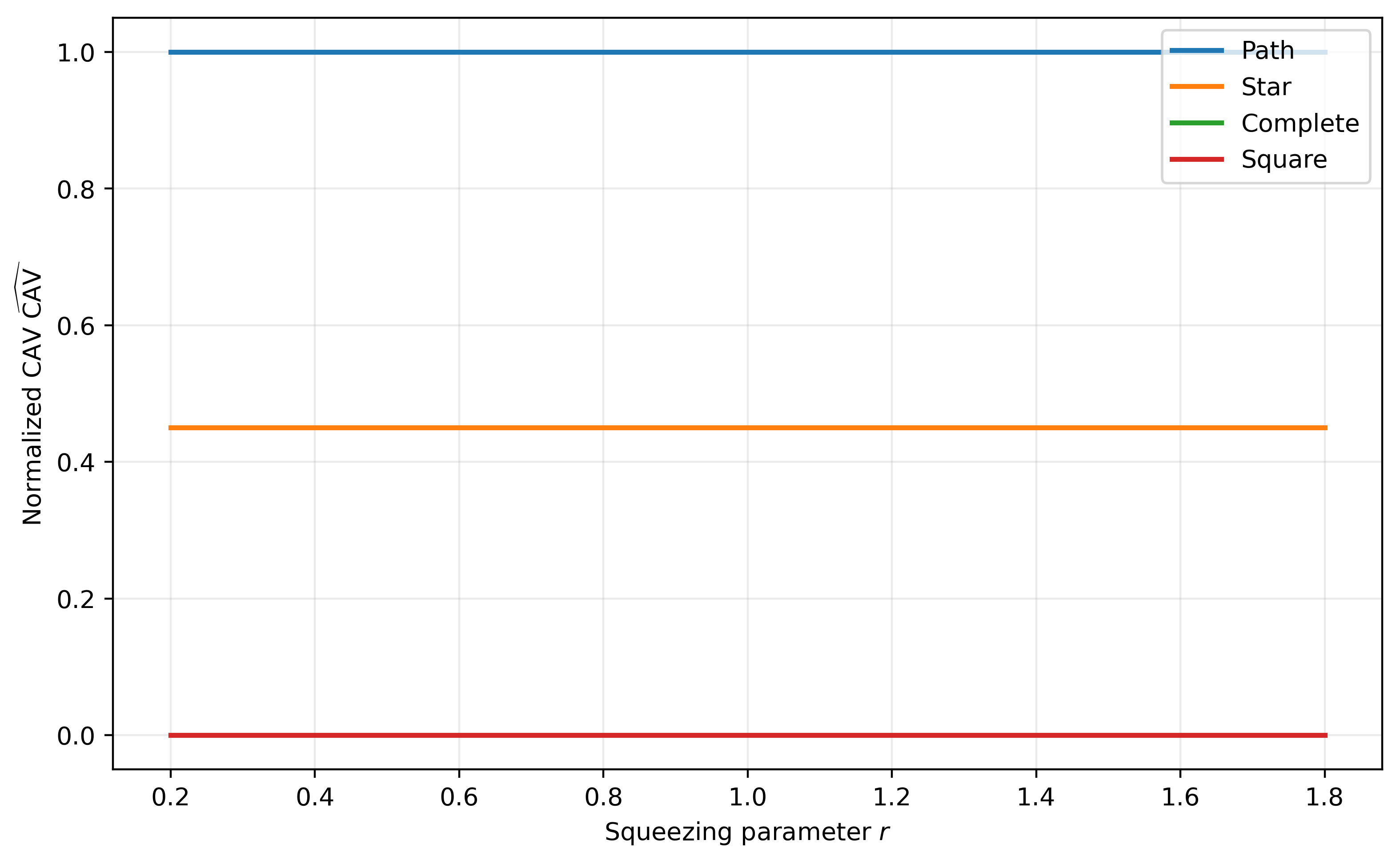}
\caption{The CAV  as a function of squeezing parameter \(r\) for the path \(P_4\), square \(C_4\), and star \(K_{1,3}\) topologies}
\label{fig:cav}
\end{figure}

\subsection{Error Vulnerability Coefficient (EVC)}
\label{sec:evc}

The stability of a continuous-variable cluster state under experimental imperfections is an important requirement for reliable measurement-based quantum computation. Practical implementations are affected by finite squeezing, optical loss, imperfect detection, phase fluctuations, and imperfections in the entangling operations, all of which perturb the covariance matrix of the Gaussian resource. Because the covariance matrix contains the complete second-order statistical description of a Gaussian state, its spectral properties provide a compact way of characterizing the conditioning of the resource.

Let $V$ denote the covariance matrix of an $N$-mode cluster state. For a positive-definite covariance matrix, its spectral condition number is
\begin{equation}
\kappa(V)=
\|V\|_2\,\|V^{-1}\|_2
=
\frac{\lambda_{\max}(V)}{\lambda_{\min}(V)},
\label{eq:evc_condition}
\end{equation}
where $\lambda_{\max}(V)$ and $\lambda_{\min}(V)$ are the largest and smallest eigenvalues of $V$, respectively. A large condition number indicates a stronger imbalance of the covariance spectrum and hence a larger sensitivity of the matrix representation to perturbations in differently scaled quadrature directions.

To obtain a bounded quantity whose favorable direction is immediately apparent, we define the Error Vulnerability Coefficient as the inverse-conditioned spectral score
\begin{equation}
\EVC(V)=\frac{1}{1+\kappa(V)}.
\label{eq:evc}
\end{equation}
Thus,
\begin{equation}
0<\EVC(V)\le\frac{1}{2},
\label{eq:evc_bounds}
\end{equation}
with larger values corresponding to a better-conditioned covariance spectrum and smaller values corresponding to stronger spectral imbalance. The use of the inverse condition number is deliberate: it converts the unbounded condition number into a bounded, monotone indicator while preserving the ordering of covariance-matrix conditioning.

For comparisons among a specified family of $N$-mode cluster states, we further remove the family-dependent scale by defining
\begin{equation}
\widehat{\EVC}(V)=
\frac{\EVC(V)-\EVC_{\min}}
{\EVC_{\max}-\EVC_{\min}},
\label{eq:evc_norm}
\end{equation}
where $\EVC_{\min}$ and $\EVC_{\max}$ are evaluated over the graph family being compared at the same system size and squeezing parameter. Consequently, $\widehat{\EVC}\in[0,1]$, where smaller values identify the less vulnerable and better-conditioned members of the comparison family, whereas larger values indicate greater spectral vulnerability.

This metric should be interpreted as a spectral robustness proxy rather than as a complete noise model. In particular, it does not replace explicit simulations of loss, detector inefficiency, phase noise, or gate errors. Instead, it provides a compact diagnostic of covariance-spectrum conditioning that can be evaluated directly from the reconstructed Gaussian covariance matrix. The distinction is important: EVC characterizes a property of the covariance representation, whereas experimentally observed fault-tolerance ultimately depends on the complete noise channel and the measurement protocol.

\begin{figure}[htbp]
\centering
\includegraphics[width=1\linewidth]{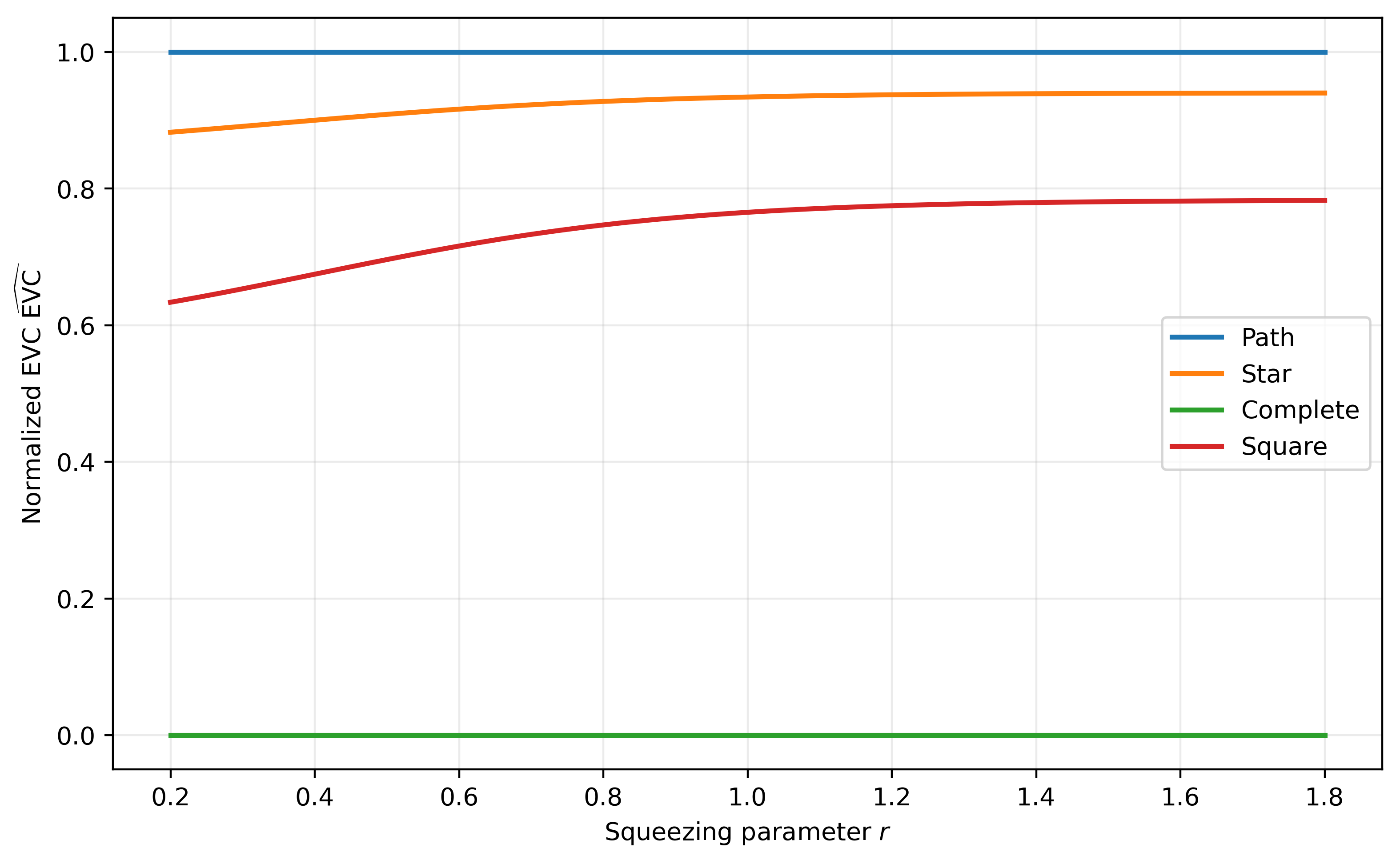}
\caption{
Normalized Error Vulnerability Coefficient (EVC) for different cluster-state topologies}
\label{fig:evc}
\end{figure}

\subsection{Communication Overhead (COM)}
\label{sec:com}

In measurement-based quantum computation, logical information propagates through the cluster state via a sequence of adaptive local measurements. Consequently, the graph-theoretic distance between two vertices provides a natural estimate of the communication effort required to transfer information across the resource. Motivated by this observation, we introduce the Communication Overhead (COM) as a graph-based metric that characterizes the average separation between pairs of modes. Cluster states with smaller COM values support shorter communication paths and are therefore expected to require fewer sequential measurement steps during computation.

Let

\begin{equation}
G=(V,E),
\qquad
N=|V|,
\label{eq:display_63}
\end{equation}

and let

\begin{equation}
d(i,j)
\label{eq:display_64}
\end{equation}

denote the shortest-path distance, measured in graph edges, between vertices $i$ and $j$. The raw communication overhead is defined as

\begin{equation}
\mathrm{COM}_{\mathrm{raw}}(G)
=
\frac{2}{N(N-1)}
\sum_{1\le i<j\le N}
d(i,j),
\label{eq:com_raw}
\end{equation}

namely the average shortest-path distance over all unordered vertex pairs.

Disconnected graphs are assigned

\begin{equation}
\mathrm{COM}_{\mathrm{raw}}(G)=\infty,
\label{eq:display_65}
\end{equation}

since information cannot be transmitted throughout the entire network, making such graphs unsuitable as universal MBQC resources.

Among connected graphs with $N$ vertices, the complete graph $K_N$ minimizes the average shortest-path distance,

\begin{equation}
\mathrm{COM}_{\mathrm{raw}}=1,
\label{eq:display_66}
\end{equation}

because every pair of vertices is directly connected. In contrast, the path graph $P_N$ maximizes the average shortest-path distance, yielding

\begin{equation}
\mathrm{COM}_{\mathrm{raw}}
=
\frac{N+1}{3}.
\label{eq:display_67}
\end{equation}

To enable direct comparison among different graph families, we normalize the communication overhead according to

\begin{equation}
\widehat{\mathrm{COM}}(G)
=
\frac{
\mathrm{COM}_{\mathrm{raw}}(G)-1
}{
\frac{N+1}{3}-1
},
\label{eq:com_norm}
\end{equation}

which maps every connected graph onto the interval

\begin{equation}
\widehat{\mathrm{COM}}
\in
[0,1].
\label{eq:display_68}
\end{equation}

With this normalization,

\begin{equation}
\widehat{\mathrm{COM}}=0
\label{eq:display_69}
\end{equation}

for the complete graph and

\begin{equation}
\widehat{\mathrm{COM}}=1
\label{eq:display_70}
\end{equation}

for the path graph. Lower values therefore correspond to shorter communication paths and more efficient information transport, whereas larger values indicate increasingly serial graph structures that require information to traverse multiple intermediate modes.

The behavior of COM also provides useful insight into the scalability of different cluster-state architectures. For a two-dimensional square lattice with linear dimension $\sqrt{N}$, the average shortest-path distance increases only as $\sqrt{N}$. Consequently,

\begin{equation}
\mathrm{COM}_{\mathrm{raw}}
=
\mathcal{O}(\sqrt{N}),
\label{eq:display_71}
\end{equation}

and the normalized metric satisfies

\begin{equation}
\widehat{\mathrm{COM}}
\rightarrow
0,
\qquad
N\rightarrow\infty.
\label{eq:display_72}
\end{equation}

This scaling highlights the favorable communication properties of lattice-based resources compared with one-dimensional graph families, whose average path length grows much more rapidly as the system size increases.

Although COM is derived entirely from graph topology, it provides a useful operational estimate of the communication depth required during measurement-based computation. Since each additional measurement stage introduces further opportunities for loss and noise accumulation, cluster states with smaller COM values are generally expected to support more efficient and more robust computational implementations. Accordingly, COM complements the correlation-based metrics developed in the previous sections by characterizing the accessibility of the available quantum resources rather than their magnitude alone.

\begin{figure}[htbp]
\centering
\includegraphics[width=1\linewidth]{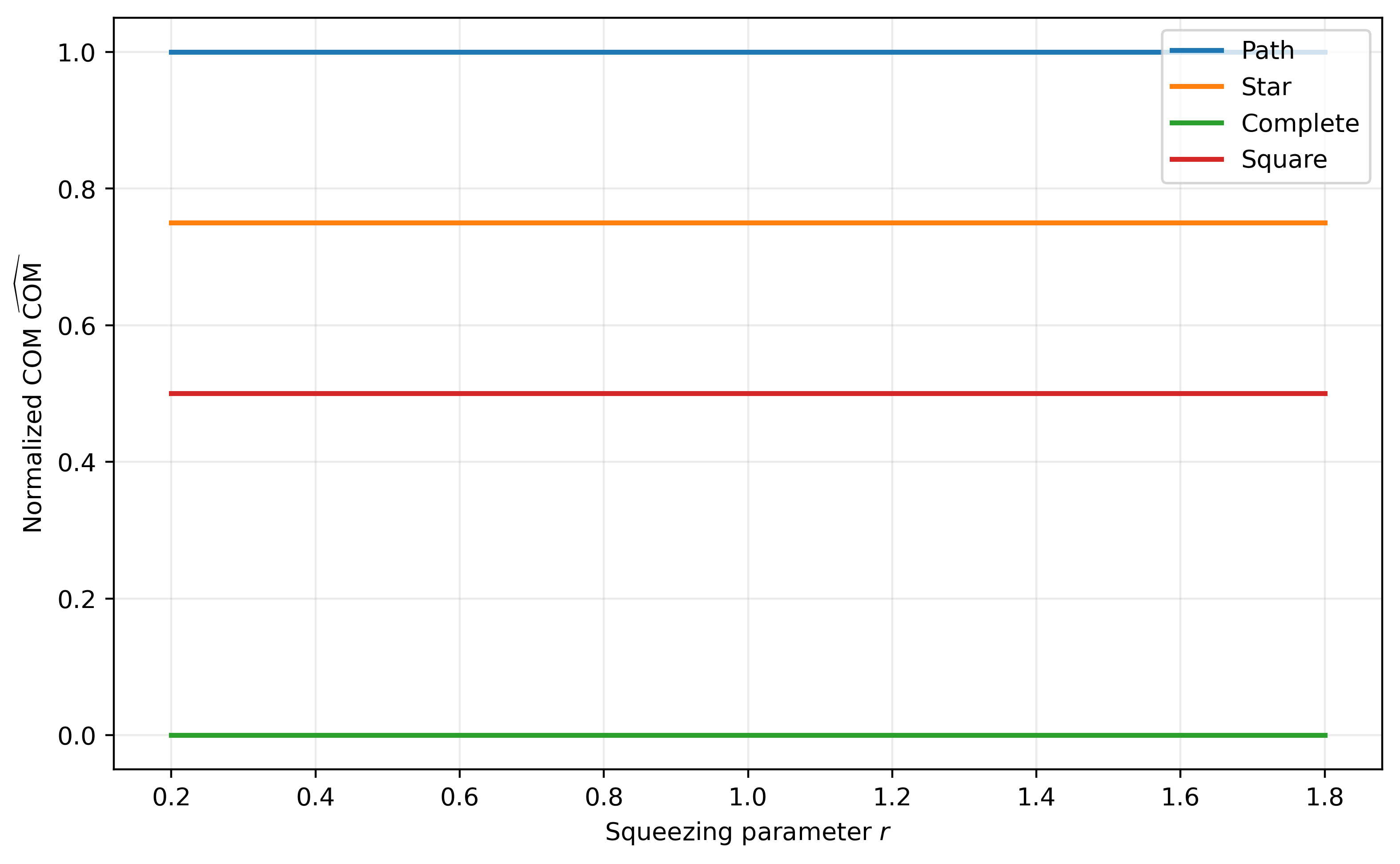}
\caption{Communication Overhead (COM) for different cluster-state topologies.}
\label{fig:com}
\end{figure}

\subsection{Redundancy Metric (RED)}
\label{sec:red}

The ability of a cluster state to support reliable quantum information processing depends not only on the existence of communication paths between modes, but also on the availability of alternative routes when part of the network becomes unavailable. Such redundancy is particularly important in measurement-based quantum computation, where imperfect measurements, mode loss, or localized hardware failures should not immediately compromise the computational resource. To quantify this structural resilience, we introduce the Redundancy Metric (RED), which measures the average availability of independent communication paths between pairs of modes.

Consider a graph

\begin{equation}
G=(V,E),
\label{eq:display_73}
\end{equation}

and let

\begin{equation}
\kappa_{ij}
\label{eq:display_74}
\end{equation}

denote the vertex connectivity between vertices $i$ and $j$, namely the minimum number of vertices whose removal disconnects the pair. By Menger's theorem, $\kappa_{ij}$ is equivalently the maximum number of internally vertex-disjoint paths connecting the same two vertices\cite{bondy2008}.

For each unordered pair of vertices, we assign the redundancy score

\begin{equation}
1-\frac{1}{\kappa_{ij}},
\label{eq:display_75}
\end{equation}

which is zero whenever only a single communication route exists and increases as additional independent paths become available. Averaging this quantity over all unordered vertex pairs yields

\begin{equation}
\mathrm{RED}(G)
=
\frac{2}{N(N-1)}
\sum_{1\le i<j\le N}
\left(
1-\frac{1}{\kappa_{ij}}
\right),
\label{eq:red_def}
\end{equation}

which naturally satisfies

\begin{equation}
\mathrm{RED}(G)\in[0,1].
\label{eq:display_76}
\end{equation}

Tree graphs, including path and star topologies, satisfy

\begin{equation}
\kappa_{ij}=1
\label{eq:display_77}
\end{equation}

for every pair of distinct vertices and therefore give

\begin{equation}
\mathrm{RED}=0,
\label{eq:display_78}
\end{equation}

reflecting the absence of alternative communication routes. At the opposite extreme, highly connected graphs possess multiple internally disjoint paths between most vertex pairs, causing RED to approach unity as the network connectivity increases. Two-dimensional lattice graphs occupy an intermediate regime in which the number of disjoint paths generally grows with system size, leading to substantially higher redundancy than in one-dimensional graph families.

Unlike several of the other metrics introduced in this work, RED is intrinsically bounded and therefore requires no additional normalization. This property makes it particularly convenient for comparing graph topologies of different sizes and connectivity patterns.

From the perspective of measurement-based quantum computation, a larger RED indicates that information can continue to propagate even after localized failures or the removal of individual modes. Consequently, RED provides a direct structural measure of fault tolerance that complements the correlation-based metrics developed earlier. While SSC, CAV, and EVC characterize the strength, distribution, and stability of the encoded correlations, RED evaluates the resilience of the underlying communication network itself.

\begin{figure}[htbp]
\centering
\includegraphics[width=1\linewidth]{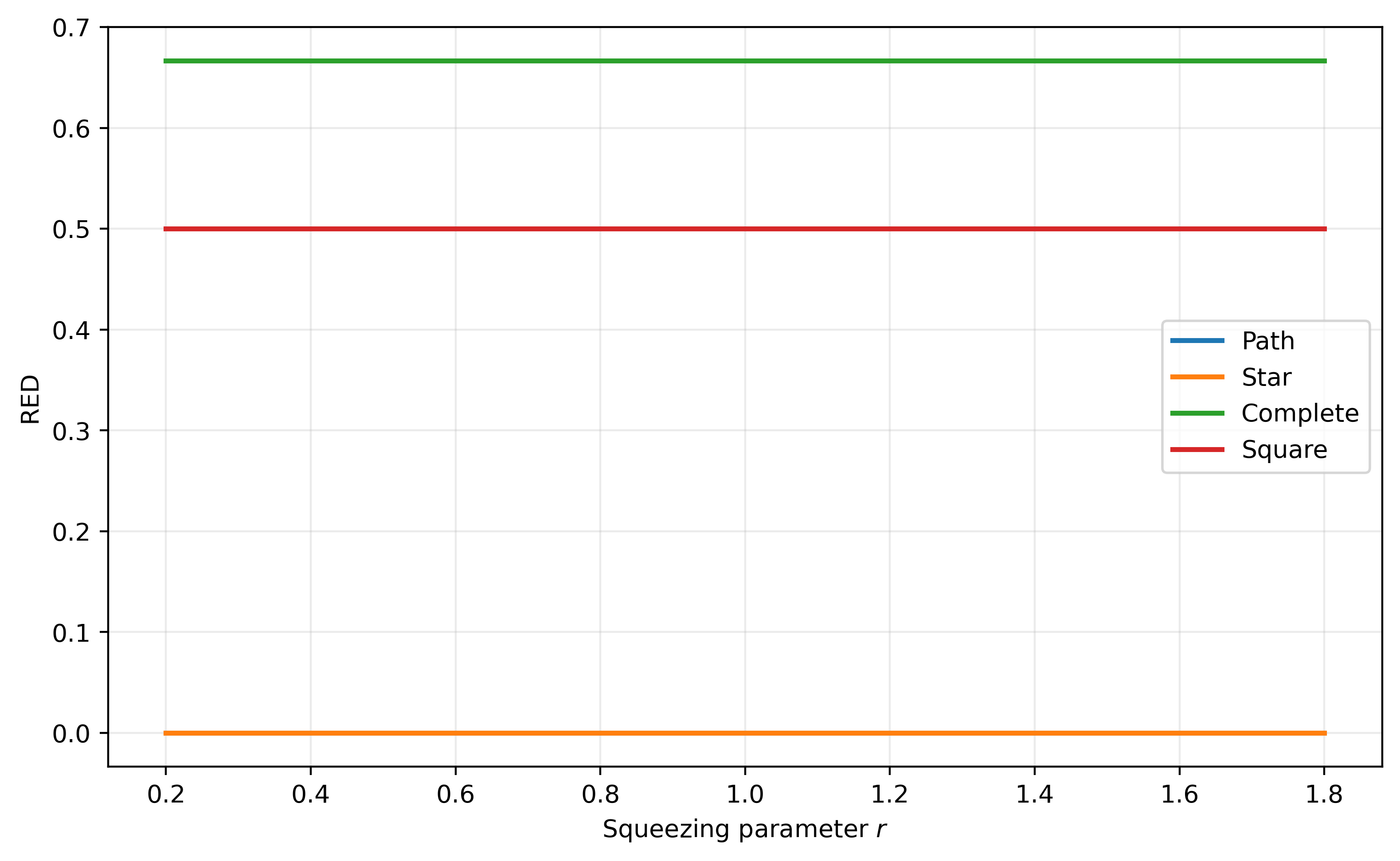}
\caption{Redundancy metric  for different cluster-state topologies.}
\label{fig:red}
\end{figure}

\subsection{Bottleneck Robustness Metric (BOT)}
\label{sec:bot}

An efficient cluster-state resource should avoid excessive reliance on a small number of highly connected modes. When the removal of a single vertex causes a substantial deterioration in the communication capability of the graph, that vertex acts as a structural bottleneck, reducing the robustness of the resource against localized failures. To quantify this effect, we introduce the Bottleneck Robustness Metric (BOT), which evaluates the largest degradation in global communication efficiency caused by the removal of any individual vertex.

For a connected graph

\begin{equation}
G=(V,E),
\qquad
|V|=N,
\label{eq:display_79}
\end{equation}

its global efficiency is defined as

\begin{equation}
E(G)
=
\frac{2}{N(N-1)}
\sum_{1\le i<j\le N}
\frac{1}{d_{ij}},
\label{eq:global_eff}
\end{equation}

where $d_{ij}$ denotes the shortest-path distance between vertices $i$ and $j$. Whenever two vertices become disconnected, the corresponding contribution is taken to be

\begin{equation}
\frac1{d_{ij}}=0.
\label{eq:display_80}
\end{equation}

For each vertex

\begin{equation}
v\in V,
\label{eq:display_81}
\end{equation}

let

\begin{equation}
G-v
\label{eq:display_82}
\end{equation}

represent the graph obtained by removing $v$ together with all incident edges. The relative loss of communication efficiency associated with removing vertex $v$ is then

\begin{equation}
B(v)
=
1-
\frac{E(G-v)}
{E(G)}.
\label{eq:display_83}
\end{equation}

The Bottleneck Robustness Metric is defined as

\begin{equation}
\mathrm{BOT}(G)
=
\max_{v\in V}
\left[
1-
\frac{E(G-v)}
{E(G)}
\right],
\label{eq:bot_def}
\end{equation}

which corresponds to the most severe degradation produced by the removal of a single vertex.

By definition,

\begin{equation}
0
\le
\mathrm{BOT}(G)
\le
1.
\label{eq:display_84}
\end{equation}

Smaller BOT values indicate that the communication efficiency is distributed relatively uniformly across the graph and is therefore insensitive to the loss of any individual mode. Conversely, larger values reveal the presence of highly critical vertices whose removal significantly degrades the communication capability of the network.

From the perspective of measurement-based quantum computation, BOT provides a direct indicator of structural robustness. Cluster states with low BOT values retain efficient communication even after localized failures because multiple regions of the graph participate in the propagation of information. In contrast, graphs exhibiting large BOT values depend strongly on one or a few central vertices, making the computational resource considerably more vulnerable to mode loss or measurement errors. Accordingly, BOT complements the RED metric by focusing on the influence of the most critical vertex, whereas RED characterizes the average availability of alternative communication paths throughout the network.
\begin{figure}[htbp]
\centering
\includegraphics[width=1\linewidth]{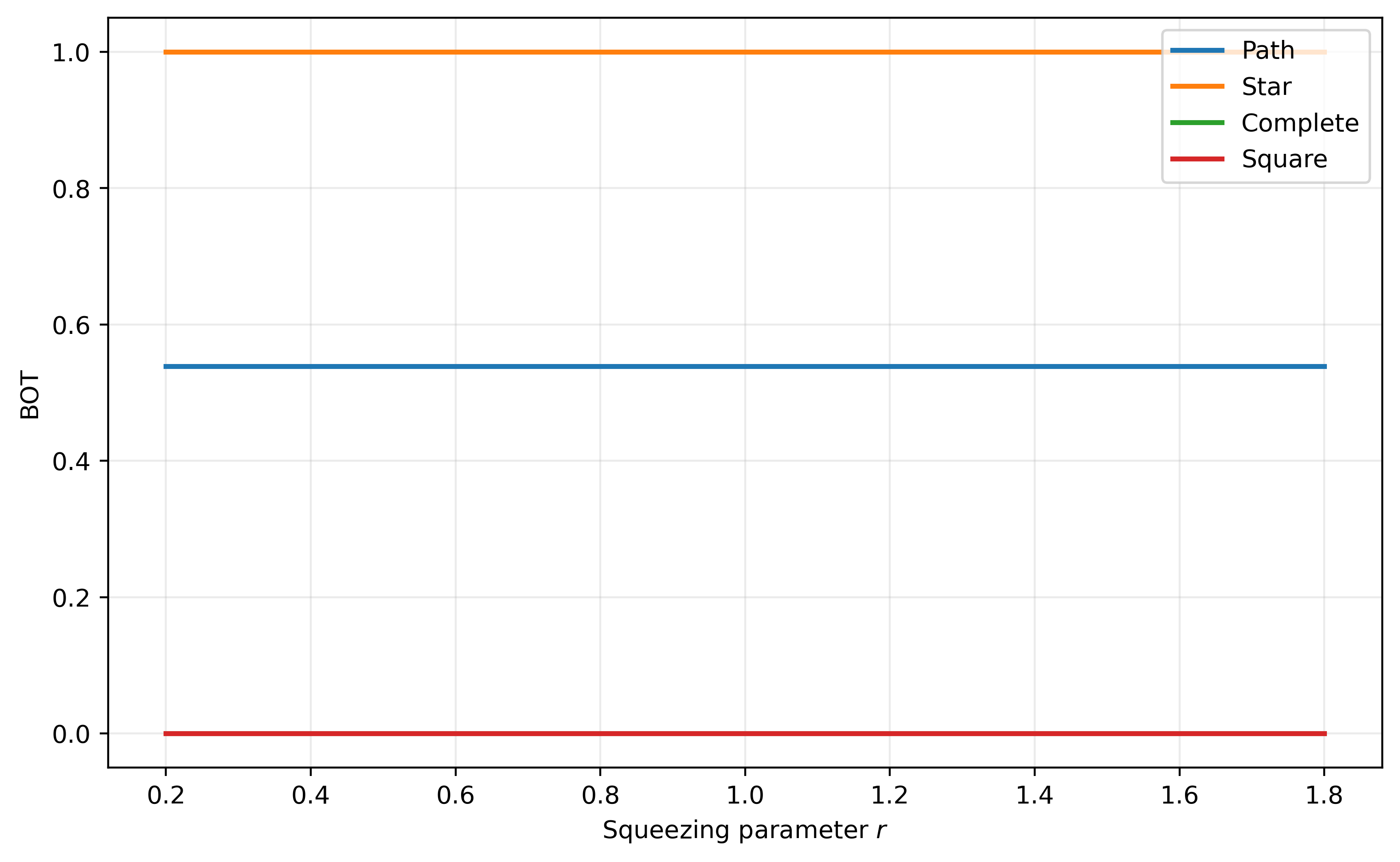}
\caption{Bottleneck vulnerability  for different cluster-state topologies.}
\label{fig:bot}
\end{figure}

\subsection{Summary of the six metrics}
\label{sec:metrics_summary}

The six metrics introduced in the preceding sections characterize complementary aspects of continuous-variable cluster states. None of them, considered individually, provides a complete assessment of the suitability of a resource for measurement-based quantum computation. Instead, each metric emphasizes a different physical or structural property, ranging from the overall strength of the encoded correlations to the topology of the communication network and its robustness against imperfections. Their combined use therefore provides a substantially more informative description than any single scalar quantity.

The six metrics naturally fall into two categories according to the direction in which their values should be optimized. Larger values of $\widehat{\SSC}$ and RED represent more desirable resources because they correspond to stronger overall correlations and greater communication redundancy, respectively. In contrast, smaller values of $\widehat{\CAV}$, $\widehat{\EVC}$, $\widehat{\mathrm{COM}}$, and $\BOT$ indicate more uniform correlation distributions, improved spectral robustness, lower communication overhead, and reduced bottleneck sensitivity.

Taken together, these metrics reveal that no individual criterion can fully characterize the quality of a cluster-state resource. A graph that performs well according to one measure may perform poorly according to another, reflecting the different physical requirements of scalable measurement-based quantum computation. This observation motivates the introduction of a unified figure of merit that simultaneously accounts for all six complementary aspects of the resource. The construction of this metric is presented in the following section.

\section{UNIFIED CLUSTER QUALITY METRIC (UCQM)}
\label{sec:ucqm}

Although each of the six metrics introduced in the previous section provides useful information about a continuous-variable cluster state, no single quantity is sufficient to characterize the overall quality of the resource. Strong correlations alone do not guarantee efficient information transfer, just as a highly connected graph may still exhibit poor robustness or an unfavorable correlation distribution. A meaningful assessment therefore requires these complementary features to be considered simultaneously. This motivates the introduction of a unified figure of merit that summarizes the overall quality of a cluster state while preserving the physical interpretation of the individual metrics.

The construction of such a metric requires one additional consideration. The six normalized diagnostics are not optimized in the same direction. Higher values of $\widehat{\SSC}$ and $\RED$ correspond to more desirable resources, whereas lower values of $\widehat{\CAV}$, $\widehat{\EVC}$, $\widehat{\mathrm{COM}}$, and $\BOT$ indicate improved performance. The unified metric is therefore designed so that every contribution consistently rewards favorable resource properties.

Let
$\widehat{\SSC}$,
$\widehat{\CAV}$,
$\widehat{\EVC}$,
$\widehat{\mathrm{COM}}$,
$\RED$,
and
$\BOT$
denote the normalized metrics introduced in Secs.~\ref{sec:ssc}--\ref{sec:bot}. We define the additive Unified Cluster Quality Metric as

\begin{equation}
\UCQM_{\mathrm{add}}(G)
=
\alpha\,\hat{\SSC}
+\gamma\,(1-\hat{\CAV})
+\delta\,(1-\hat{\EVC})
+\epsilon\,(1-\widehat{\mathrm{COM}})
+\zeta\,\RED
+\eta\,(1-\BOT),
\label{eq:ucqm_add}
\end{equation}
where the non-negative coefficients satisfy
\begin{equation}
\alpha+\gamma+\delta+\epsilon+\zeta+\eta=1.
\label{eq:ucqm_weight_sum}
\end{equation}
Because all six quantities lie in the interval $[0,1]$, the additive UCQM is itself bounded between zero and one. The terms
$(1-\widehat{\CAV})$,
$(1-\widehat{\EVC})$,
$(1-\widehat{\mathrm{COM}})$,
and
$(1-\BOT)$
simply reverse the optimization direction of those metrics for which smaller values represent better performance. Consequently, larger values of $\UCQM_{\mathrm{add}}$ always correspond to higher overall cluster-state quality.

To illustrate the proposed framework, the numerical examples presented in this work employ the default weighting scheme summarized in Table~\ref{tab:ucqm_weights}. These coefficients are intended as a practical baseline rather than a unique or optimal choice. Greater emphasis is assigned to the total correlation strength and to robustness-related properties, reflecting their central role in scalable measurement-based quantum computation. Depending on the target application or experimental platform, different weighting schemes may be adopted without modifying the structure of the UCQM itself.

\begin{table}[htbp]
\centering
\caption{Default weights used in the additive UCQM.}
\label{tab:ucqm_weights}
\renewcommand{\arraystretch}{1.15}
\begin{tabular}{lcc}
\toprule
\textbf{Metric} & \textbf{Symbol} & \textbf{Weight} \\
\midrule
Total correlation strength & $\hat{\SSC}$ & $\alpha = 0.40$ \\
Anisotropy & $1-\hat{\CAV}$ & $\gamma = 0.10$ \\
Error robustness & $1-\hat{\EVC}$ & $\delta = 0.15$ \\
Communication efficiency & $1-\widehat{\mathrm{COM}}$ & $\epsilon = 0.10$ \\
Redundancy & $\RED$ & $\zeta = 0.15$ \\
Bottleneck robustness & $1-\BOT$ & $\eta = 0.10$ \\
\bottomrule
\end{tabular}
\end{table}

A multiplicative alternative is also natural when one wishes to enforce a stricter notion of quality, in which a single weak ingredient can substantially suppress the overall score. In that case one may define
\begin{equation}
\UCQM_{\times}(G)
=
\hat{\SSC}\,(1-\hat{\CAV})\,(1-\hat{\EVC})\,(1-\widehat{\mathrm{COM}})\,\mathrm{RED}\,(1-\mathrm{BOT}).
\label{eq:ucqm_mult}
\end{equation}

This form also lies in $[0,1]$, but unlike the additive version it vanishes whenever any factor vanishes. It is therefore more severe and is best interpreted as a joint feasibility indicator. In the present work, we prefer the additive form, since it reflects the fact that realistic cluster states typically exhibit gradual trade-offs rather than binary success or failure across all six criteria.

\begin{figure}[htbp]
\centering
\includegraphics[width=0.85\linewidth]{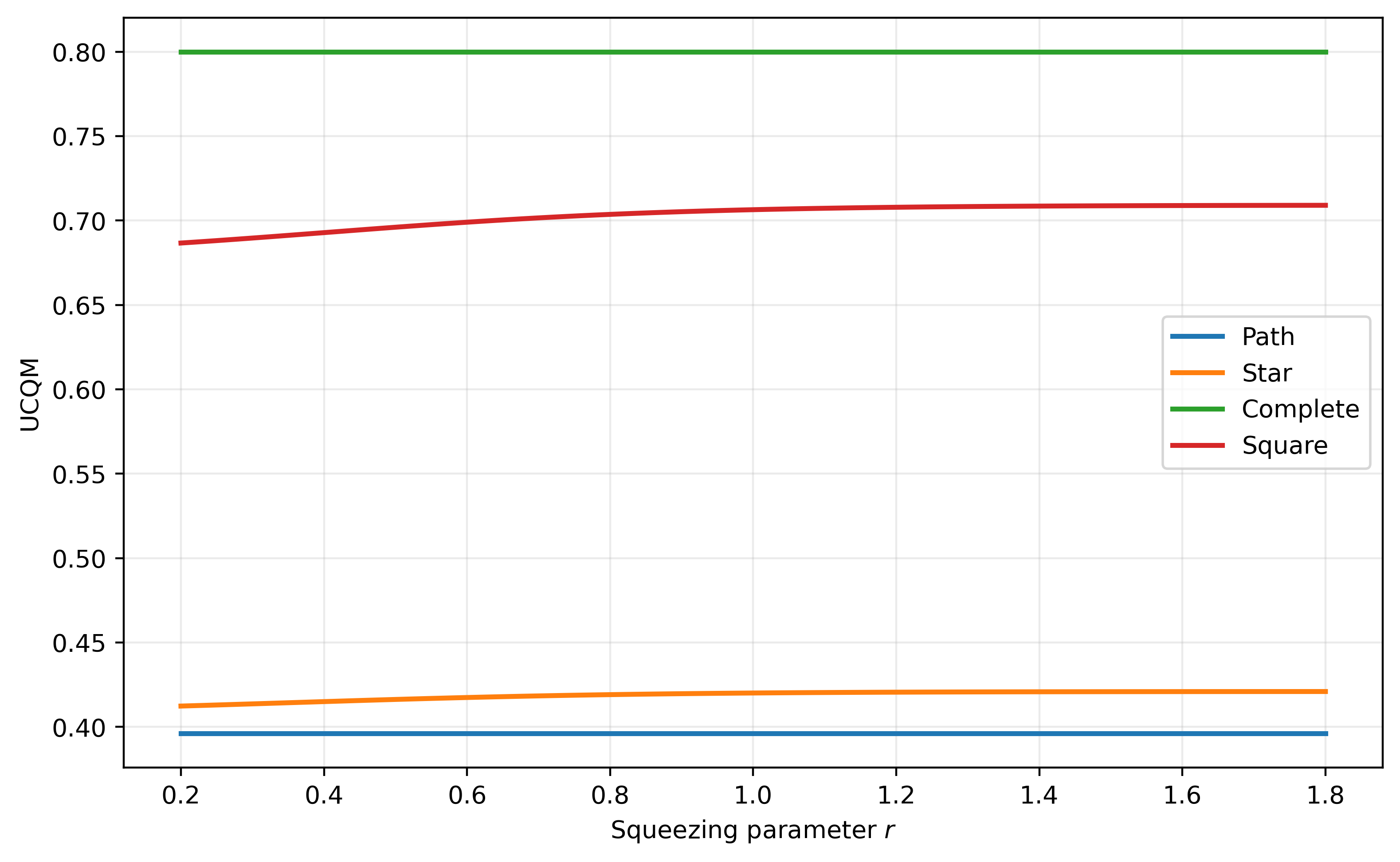}
\caption{Unified cluster quality metric as a function of squeezing parameter $r$ for representative topologies. The square lattice remains the strongest performer across the full range, reflecting its favorable balance of correlation strength, redundancy, and robustness.}
\label{fig:ucqm}
\end{figure}

The UCQM provides a convenient basis for interpreting numerical results. In the parameter range explored here, larger values indicate more favorable resources, whereas lower values signal that one or more essential properties are insufficiently developed. Since the score is normalized, it enables direct comparison between graph families and squeezing levels within the same framework. For the canonical families studied in this work, the square topology and its larger two-dimensional extensions consistently achieve the highest scores, whereas path and star graphs perform significantly worse, especially as the system size grows. This trend is reflected in Fig.~\ref{fig:ucqm}.

\begin{table}[htbp]
\centering
\caption{Representative UCQM scores for canonical topologies using the additive form and the default weights.}
\label{tab:ucqm_examples}
\renewcommand{\arraystretch}{1.15}
\begin{tabular}{lcc}
\toprule
\textbf{Topology} & $N=4$ & $N=100$ \\
\midrule
Square / 2D lattice & $0.70$ & $0.40$ \\
Path & $0.39$ & $0.30$ \\
Star & $0.41$ & $0.23$ \\
\bottomrule
\end{tabular}
\end{table}

The present framework is intentionally restricted to Gaussian states with uniform squeezing and idealized CZ entangling gates. Extensions to non-Gaussian resources, such as GKP-type encodings, would require incorporating higher-order statistical information beyond the covariance matrix. Likewise, inhomogeneous squeezing would call for mode-dependent weighting of the graph structure, and more realistic models of loss, detector inefficiency, or phase noise would require explicit channel-dependent corrections. Disconnected graphs are assigned $\UCQM=0$ by definition, since they cannot support computation across the full network. These refinements are important, but they lie beyond the scope of the present work.

\section{Scalability and Computational Complexity}
\label{sec:scalability}

Beyond physical relevance, a practical assessment framework must also remain computationally tractable as the size of the cluster state increases. Since realistic measurement-based quantum computation ultimately targets large-scale optical networks, the computational effort required to evaluate the proposed metrics becomes an important consideration. The objective is therefore not only to characterize cluster states accurately, but also to ensure that the evaluation procedure itself scales efficiently with the number of modes.

Within the present framework, every quantity is determined by two ingredients: the graph describing the cluster-state connectivity and the corresponding covariance matrix. Consequently, the computational complexity of each metric is governed by the amount of graph-theoretic or spectral information that must be extracted from these two objects.

The least demanding quantities are the correlation-based metrics, namely $\SSC$ and $\CAV$. Both are obtained directly from covariance-matrix elements and require only algebraic operations on matrix entries. When the covariance matrix is stored explicitly, their computational cost is at most quadratic in the number of modes. For sparse cluster graphs, the effective cost is often considerably lower because only the nonzero correlations associated with existing edges need to be processed.

The Error Vulnerability Coefficient is computationally more expensive because it depends on the spectral properties of the covariance matrix through its condition number. A straightforward evaluation requires diagonalization of the full $2N\times2N$ covariance matrix, leading to a worst-case complexity of $\Theta(N^3)$\cite{slapnicar2013symmetric}. In practice, however, structured covariance matrices and iterative eigensolvers can substantially reduce the computational effort for large sparse systems.

The remaining metrics rely primarily on graph-theoretic operations. Computing COM requires shortest-path distances between vertex pairs, RED depends on evaluating internally vertex-disjoint paths, while BOT measures the change in global efficiency after removing individual vertices. Although these calculations are generally more involved than simple covariance-based quantities, efficient algorithms are available for sparse and structured graphs commonly encountered in measurement-based quantum computation.

Taken together, these observations imply that the overall computational cost of the framework is dominated by the spectral decomposition required for EVC together with the graph algorithms underlying COM, RED, and BOT. Consequently, the complete evaluation of all six metrics remains bounded by $\Theta(N^3)$ in the worst case. The corresponding memory requirement is $\Theta(N^2)$ when either the covariance matrix or the full distance matrix is stored explicitly, although sparse matrix representations can reduce this requirement significantly for low-degree graph families.

The asymptotic behavior of the metrics also depends strongly on the topology of the underlying graph. For the path graph $P_N$, the number of edges grows linearly with the number of vertices, $|E|=N-1$, implying that the normalized correlation strength satisfies $\widehat{\SSC}=\Theta(1/N)$. The same asymptotic scaling is obtained for two-dimensional square lattices because the total number of nearest-neighbor edges also increases linearly with $N$, despite the substantially different connectivity pattern. Star graphs exhibit an identical edge-density scaling, yet their highly centralized topology results in poor redundancy and pronounced bottleneck vulnerability. At the opposite extreme, the complete graph achieves the maximal normalized correlation strength, $\widehat{\SSC}=1$, although its quadratic growth in the number of edges makes it unsuitable as a realistic architecture for large-scale implementations.

The asymptotic behavior of the principal metrics for the canonical graph families considered throughout this work is summarized in Table~\ref{tab:scaling_summary}. Rather than emphasizing numerical prefactors, the table highlights the dominant scaling trends that ultimately determine whether each resource remains useful as the system size increases. From this perspective, two-dimensional lattice architectures exhibit the most balanced behavior, maintaining favorable communication efficiency, redundancy, and robustness without incurring excessive structural complexity.

\section{Experimental Considerations}
\label{sec:experimental}

The proposed framework is formulated so that its required inputs can, in principle, be obtained from experimentally reconstructed covariance matrices together with the experimentally realized or intended connectivity graph. For Gaussian optical states, homodyne detection provides direct access to quadrature statistics and therefore to the first and second moments required to reconstruct the covariance matrix \cite{lvovsky2009homodyne}. The present metrics use only second-order information from the Gaussian resource, together with graph-theoretic quantities that can be determined from the mode-connectivity pattern.

In an experimental implementation, the reconstructed covariance matrix $V_{\mathrm{exp}}$ can be inserted directly into the state-dependent metrics. In particular, SSC is evaluated from the measured cross-quadrature covariance elements, CAV from the pairwise weights constructed from the measured covariance structure and the corresponding adjacency matrix, and EVC from the eigenvalue spectrum of $V_{\mathrm{exp}}$. The graph-dependent quantities COM, RED, and BOT can then be evaluated from the experimentally realized connectivity graph. This separation is useful because it allows deviations from the ideal covariance model to be quantified without requiring the experimental state to be replaced by an ideal theoretical covariance matrix.

For a measured covariance matrix, finite sampling, calibration uncertainty, detector inefficiency, optical loss, and mode mismatch can introduce statistical and systematic errors. Accordingly, experimentally reported values of the six metrics should preferably be accompanied by uncertainty estimates obtained by resampling the measured data or by propagating the covariance-matrix reconstruction errors. The sample-variance form used in CAV is particularly compatible with such resampling procedures, while the spectral EVC should be evaluated with care when the smallest covariance eigenvalue is close to the experimental uncertainty floor.

The present framework does not assume that the experimentally reconstructed state is perfectly described by the ideal covariance matrix of Eq.~\eqref{eq:cov_final}. Instead, Eq.~\eqref{eq:cov_final} defines the reference model against which experimentally reconstructed states can be benchmarked. This distinction is important for future applications: replacing the ideal covariance matrix by $V_{\mathrm{exp}}$ allows the same metric definitions to remain applicable in the presence of realistic preparation imperfections, while explicit channel models can be incorporated when a detailed noise analysis is required.

\section{Numerical Results}
\label{sec:results}

Having established the analytical definitions of the six metrics, we now examine their behavior for representative continuous-variable cluster-state topologies. Unless stated otherwise, the covariance matrices are generated from Eq.~\eqref{eq:cov_final} under the assumption of uniform squeezing, with the squeezing parameter varied over the interval
\begin{equation}
0.2\le r\le1.8.
\label{eq:display_85}
\end{equation}
Representative numerical values are reported for $r=0.80$. Throughout this section, all six metrics are evaluated directly from the covariance matrix and the corresponding graph, while the overall cluster quality is quantified using the additive UCQM introduced in Sec.~\ref{sec:ucqm}.


\subsection{Four-mode benchmark graphs}

We begin with the smallest nontrivial examples, namely the three canonical four-mode graphs summarized in Table~\ref{tab:fourmode}. Despite their identical system size, these topologies exhibit markedly different structural characteristics, making them a convenient benchmark for assessing the discriminative power of the proposed metrics.
\label{tab:six_interpretation}
\begin{table}[htbp]
\centering
\caption{Six normalized metrics and the resulting UCQM for representative four-mode cluster-state topologies at $r=0.80$.}
\label{tab:fourmode}
\renewcommand{\arraystretch}{1.15}
\begin{tabular}{lccccccc}
\toprule
& $\hat{\SSC}$ & $\hat{\CAV}$ & $\hat{\EVC}$ & $\widehat{\mathrm{COM}}$ & $\RED$ & $\BOT$ & $\UCQM$ \\
\midrule
$P_4$ (path) & 0.50 & 1.00 & 1.00 & 1.00 & 0.00 & 0.53 & 0.39 \\
$C_4$ (square) & 0.66 & 0.00 & 0.74 & 0.50 & 0.50 & 0.0 & 0.70 \\
$K_{1,3}$ (star) & 0.50 & 0.45 & 0.92 & 0.75 & 0.00 & 1.00 & 0.41 \\
\bottomrule
\end{tabular}
\end{table}

Among these three configurations, the square graph achieves the highest overall UCQM. Its advantage does not originate from a single outstanding property but from a balanced combination of strong correlation strength, relatively uniform correlation distribution, moderate communication overhead, and the presence of redundant communication paths. In contrast, the path graph performs reasonably well with respect to total correlation strength but suffers from its inherently one-dimensional geometry, which limits communication efficiency and completely eliminates path redundancy. The star topology represents the opposite extreme: although every peripheral mode is connected to the central hub, this highly centralized architecture produces maximal correlation anisotropy together with the largest bottleneck vulnerability, ultimately yielding the lowest overall quality score.

These observations illustrate an important point. The amount of correlation contained in a cluster state is only one ingredient of a useful computational resource. Equally important is the manner in which those correlations are distributed throughout the graph, since distributed connectivity ultimately determines whether information can propagate efficiently and robustly during measurement-based quantum computation.


\subsection{Scaling to larger graphs}

The same qualitative behavior persists as the system size increases. Representative results for graphs containing one hundred modes are summarized in Table~\ref{tab:largescale}. Although all three graph families become substantially larger, their relative ordering remains unchanged.

\begin{table}[htbp]
\centering
\caption{Representative normalized metric values for canonical $N=100$ cluster-state topologies evaluated at $r=0.80$.}
\label{tab:largescale}
\renewcommand{\arraystretch}{1.15}
\begin{tabular}{lccccccc}
\toprule
& $\hat{\SSC}$ & $\hat{\CAV}$ & $\hat{\EVC}$ & $\widehat{\mathrm{COM}}$ & $\RED$ & $\BOT$ & $\UCQM$ \\
\midrule
$P_{100}$ (path) & 0.02 & 0.34 & 1.00 & 1.00 & 0.00 & 0.15 & 0.30 \\
$10\times10$ lattice & 0.03 & 1.00 & 0.73 & 0.17 & 0.68 & 0.00 & 0.40 \\
$K_{1,99}$ (star) & 0.02 & 0.07 & 0.23 & 0.03 & 0.00 & 1.00 & 0.23 \\
\bottomrule
\end{tabular}
\end{table}

The square lattice continues to exhibit the most balanced performance as the graph grows, preserving favorable communication efficiency while maintaining both high redundancy and strong resistance to bottleneck failures. By comparison, the path graph becomes increasingly communication limited because the average shortest-path distance increases steadily with system size, whereas no alternative communication routes emerge. The star graph remains dominated by its central hub, making the resource progressively more susceptible to localized failures despite its short average path length.

An important conclusion follows from these numerical comparisons. Conventional entanglement measures, including logarithmic negativity and von Neumann entropy, are capable of quantifying the overall amount of quantum correlation, yet they provide little information about how that correlation is organized within the computational resource. Consequently, states exhibiting comparable global entanglement may differ substantially in their practical suitability for measurement-based quantum computation.

By combining correlation strength, correlation distribution, robustness against perturbations, communication efficiency, structural redundancy, and bottleneck sensitivity within a single normalized framework, the UCQM resolves this ambiguity. The numerical results consistently identify two-dimensional lattice architectures as the most balanced resources among the canonical graph families considered here, supporting their widely recognized role as scalable platforms for continuous-variable measurement-based quantum computation.

\section{Toward Fault-Tolerant CV MBQC}
\label{sec:fault_tolerance}

Although the UCQM framework has been developed for ideal Gaussian cluster states, its scope naturally extends toward fault-tolerant continuous-variable measurement-based quantum computation. In this setting, resource quality is determined not only by the graph topology itself, but also by how effectively the generated cluster suppresses finite-squeezing noise and supports logical error correction. This connection is most naturally expressed through the nullifier formalism, which provides the standard description of continuous-variable cluster states in the presence of realistic imperfections \cite{menicucci2011temporal, walshe2019continuous}.

For an ideal CV cluster state associated with an adjacency matrix $A$, the nullifier operators are given by

\begin{equation}
\hat{\delta}_i=\hat{p}_i-\sum_{j=1}^{N}A_{ij}\hat{x}_j,
\label{eq:nullifier}
\end{equation}

and vanish only in the infinite-squeezing limit. Practical implementations inevitably operate with finite squeezing, resulting in nonzero nullifier variances. These residual fluctuations quantify the departure of the prepared state from the ideal cluster-state manifold and therefore determine the amount of excess Gaussian noise introduced during measurement-based computation \cite{walshe2019continuous}. Their importance becomes even more pronounced in architectures employing Gottesman--Kitaev--Preskill (GKP) encoding, where logical qubits are embedded into the Gaussian resource and the attainable logical performance depends directly on the underlying physical noise budget \cite{gottesman2001encoding,Baragiola2019,vuillot2023fault}.

Viewed from this perspective, the six metrics introduced in the previous sections acquire a more operational interpretation. The normalized total correlation strength $\hat{\SSC}$ reflects how faithfully the physical resource realizes the intended graph-induced correlations. The anisotropy measure $\hat{\CAV}$ characterizes the spatial uniformity of these correlations; excessive anisotropy may produce preferential directions for noise propagation and therefore lead to unequal logical performance across the computational graph \cite{walshe2019continuous, alexander2016scalable}. Likewise, the Error Vulnerability Coefficient $\hat{\EVC}$ provides a bounded inverse-conditioning indicator of the covariance spectrum. Larger values correspond to better conditioning and therefore to a reduced sensitivity of the covariance representation to perturbations.

The graph-theoretic quantities $\RED$ and $\BOT$ become particularly significant once fault tolerance is considered. Reliable logical computation requires that both encoded quantum information and adaptive classical feed-forward signals remain robust against localized failures. High redundancy offers multiple independent propagation routes, reducing the impact of individual defective modes, while a low bottleneck score indicates that no single vertex dominates the global communication structure. Together, these properties enhance the resilience of the computational resource against photon loss, imperfect measurements, and other localized imperfections that inevitably arise in photonic implementations \cite{menicucci2011temporal, vuillot2023fault}.

These considerations motivate an extended quality functional that supplements UCQM with quantities directly associated with fault-tolerant performance. One possible generalization is

\begin{equation}
\UCQM_{\mathrm{FT}}
=
\sum_{k=1}^{6}w_kM_k
+
w_7\mathcal{R}_{\mathrm N}
+
w_8\mathcal{R}_{\mathrm{GKP}}
+
w_9\mathcal{R}_{\mathrm{th}},
\label{eq:ucqm_ft}
\end{equation}

where $M_k$ denotes the six normalized metrics introduced earlier, $\mathcal{R}_{\mathrm N}$ measures robustness against nullifier fluctuations, $\mathcal{R}_{\mathrm{GKP}}$ characterizes compatibility with GKP logical encoding, and $\mathcal{R}_{\mathrm{th}}$ quantifies proximity to a desired fault-tolerance threshold. A convenient choice for the nullifier contribution is

\begin{equation}
\mathcal{R}_{\mathrm N}
=
\left[
1+
\frac{1}{N}
\sum_{i=1}^{N}
\Delta^2(\hat{\delta}_i)
\right]^{-1},
\label{eq:r_nullifier}
\end{equation}

which decreases monotonically as the average nullifier variance increases. Other monotonic functions could be adopted when different experimental noise models are considered. The weighting coefficients satisfy $w_i\ge0$ together with the normalization condition $\sum_{i=1}^{9}w_i=1$.

The principal advantage of this extension is that it unifies architectural quality with logical robustness within a single quantitative framework. A cluster state exhibiting strong correlations but poor redundancy may still perform unsatisfactorily once realistic noise is taken into account, since errors tend to accumulate along a small number of critical communication channels. Conversely, graph structures that distribute connectivity more evenly suppress the concentration of nullifier noise and support more stable logical information flow. Consequently, the UCQM framework can naturally evolve from a purely structural benchmarking tool into a noise-aware metric for evaluating realistic continuous-variable computational resources.

Such an extension is particularly relevant for emerging large-scale photonic platforms, where finite squeezing and optical loss remain the principal limitations to fault-tolerant operation \cite{asavanant2019generation, larsen2021fault}. In these architectures, maximizing entanglement alone is insufficient; what ultimately matters is the amount of useful entanglement that remains operational under realistic experimental conditions. By incorporating nullifier quality together with graph-theoretic robustness into a unified score, the extended UCQM establishes a direct connection between resource-state design and the practical requirements of scalable fault-tolerant continuous-variable quantum computation.

\section{Discussion}
\label{sec:discussion}

The results obtained throughout this work reveal a consistent picture of how graph topology influences the usefulness of continuous-variable cluster states for measurement-based quantum computation. Rather than identifying a single dominant property, the proposed framework shows that high-quality computational resources emerge from a balance between several complementary characteristics. Correlation strength, correlation distribution, communication efficiency, structural redundancy, and robustness against perturbations all contribute to the overall performance, and no individual metric is sufficient to characterize the resource by itself.

Among the canonical graph families investigated here, the two-dimensional square lattice consistently exhibits the most favorable balance. Although its total correlation strength is not dramatically different from that of other sparse graphs under the adopted normalization, its correlations remain distributed much more uniformly across the network. At the same time, its covariance spectrum stays well conditioned, communication paths increase only with the linear dimension of the lattice, and the graph naturally provides multiple alternative routes between distant modes. Equally important, no single vertex dominates the information flow, making the lattice substantially less sensitive to localized failures than hub-based architectures. These complementary features explain why the lattice achieves the highest UCQM over the entire range of system sizes considered in this study.

The remaining graph families illustrate different forms of structural imbalance. Path graphs possess only minimal connectivity, so information must propagate sequentially through long chains of intermediate modes. As the number of modes increases, the average communication distance grows steadily while redundancy remains absent, causing their overall quality to deteriorate. Star graphs exhibit the opposite behavior. Their central vertex provides efficient short-range communication, but at the cost of creating a pronounced structural bottleneck. The entire resource becomes highly dependent on a single mode, making the graph particularly vulnerable to localized imperfections or mode loss. The complete graph represents the opposite extreme by maximizing connectivity and minimizing communication distance, yet its dense interaction pattern makes it unrealistic for scalable experimental preparation and therefore more suitable as an ideal reference than as a practical architecture.

These structural differences become particularly transparent when the six proposed metrics are considered together. Each metric captures a distinct physical aspect of the resource, and their preferred optimization directions naturally differ. Larger values of $\hat{\SSC}$ and $\RED$ indicate stronger global correlations and a richer set of alternative communication paths, respectively. In contrast, smaller values of $\hat{\CAV}$, $\widehat{\mathrm{COM}}$, and $\BOT$, $\hat{\EVC}$ correspond to a more homogeneous correlation landscape, better covariance-spectrum conditioning, shorter communication paths, and reduced dependence on critical vertices. The advantage of the proposed framework therefore lies not in replacing existing indicators with another single scalar quantity, but in combining complementary structural information into a coherent and physically interpretable assessment.

An important implication of these results is that conventional entanglement measures alone do not fully characterize the computational usefulness of a cluster state. Quantities such as logarithmic negativity or von Neumann entropy successfully quantify the amount of quantum correlation present in a state, yet they provide little information about how those correlations are spatially organized or how efficiently they can support measurement-based computation. Two cluster states with comparable entanglement may therefore exhibit markedly different computational performance because their underlying graph structures differ. By explicitly incorporating topological organization, communication efficiency, redundancy, and robustness, the UCQM framework distinguishes between resources that would otherwise appear similar from the perspective of traditional entanglement measures.

The present analysis is intentionally restricted to Gaussian cluster states generated from uniformly squeezed modes through ideal controlled-phase interactions. This simplified setting allows the proposed metrics to be expressed directly in terms of covariance matrices and graph-theoretic quantities while keeping the resulting framework computationally tractable. Nevertheless, several natural extensions deserve further investigation. Incorporating mode-dependent squeezing, realistic loss channels, detector inefficiency, phase fluctuations, or other hardware-specific imperfections would provide a more detailed description of experimentally generated resources. Likewise, extending the framework to non-Gaussian cluster states, including GKP-encoded or cat-state architectures, will require additional descriptors beyond second-order moments. Such extensions are expected to refine rather than alter the central idea developed here: evaluating cluster-state resources requires simultaneously accounting for both their quantum correlations and the graph structures through which those correlations support computation.

More broadly, the proposed framework offers a systematic methodology for comparing cluster-state architectures across different physical platforms. Because each metric has a clear operational interpretation and can be evaluated directly from experimentally accessible covariance matrices and graph descriptions, the framework can serve both as a theoretical benchmarking tool and as a practical design guideline for future large-scale implementations of continuous-variable measurement-based quantum computation.

At the same time, it is important to be clear about the scope of the present framework. The analysis assumes Gaussian cluster states, uniform squeezing, and ideal CZ gates, and therefore does not capture all of the complications that arise in realistic experiments. Non-Gaussian resources, such as GKP-type encodings or cat-state variants, would require additional structure beyond the covariance matrix, while mode-dependent squeezing would naturally call for a vertex-weighted extension of the graph description. Likewise, correlated loss, phase noise, and hardware-specific imperfections are not yet included explicitly. These are meaningful directions for future work, and they would likely enrich the framework rather than invalidate it. In particular, tuning the metric weights to a specific experimental platform, possibly with data-driven methods, appears to be a promising way to adapt the general formalism to concrete devices.

\section{Conclusion}
\label{sec:conclusion}

This work has presented a unified framework for the quantitative assessment of continuous-variable cluster states through six complementary metrics that characterize distinct physical aspects of the resource. The proposed diagnostics quantify total correlation strength (SSC), correlation anisotropy (CAV), spectral sensitivity to perturbations (EVC), communication overhead (COM), path redundancy (RED), and bottleneck vulnerability (BOT). By combining these quantities into the Unified Cluster Quality Metric (UCQM), the framework provides a single normalized score while preserving the physical interpretation of the individual components.

The numerical analysis demonstrates that different graph topologies exhibit markedly different resource characteristics, even when they contain comparable levels of overall correlation. Among the canonical families investigated here, the two-dimensional square lattice consistently achieves the most balanced performance. It combines substantial redundancy with low bottleneck vulnerability, moderate communication overhead, and bounded sensitivity to perturbations, allowing its overall quality to remain favorable as the system size increases. In contrast, path and star graphs each suffer from inherent structural limitations that become increasingly pronounced in larger systems, while the complete graph serves primarily as an idealized reference rather than a realistic architecture for scalable implementation.

An important feature of the proposed framework is its computational practicality. For dense graphs, the overall computational cost is dominated by spectral and graph-theoretic operations and scales as $\Theta(N^3)$, whereas sparse and structured topologies admit significantly more efficient evaluation strategies. Since all required quantities can be extracted from covariance matrices reconstructed through standard homodyne tomography, the framework is directly applicable to current experimental continuous-variable photonic platforms without requiring full quantum-state tomography beyond the Gaussian description.

More broadly, the UCQM framework provides a systematic methodology for comparing cluster-state architectures from multiple complementary perspectives instead of relying on a single entanglement-based indicator. Although the present study has focused on Gaussian cluster states generated with uniform squeezing and ideal CZ interactions, the formalism naturally admits future extensions that incorporate non-Gaussian resources, mode-dependent squeezing, realistic noise channels, and fault-tolerant logical encodings. We expect that such developments will further strengthen the role of UCQM as a practical benchmarking tool for the design, optimization, and experimental evaluation of scalable continuous-variable measurement-based quantum computing architectures.

\section*{Author Contributions}

S.S. conceived the study, developed the theoretical framework, formulated the  metrics, performed the analytical and numerical analysis, and wrote the manuscript.

\section*{Funding}

This research received no specific grant from any funding agency in the public, commercial, or not-for-profit sectors.

\section*{Data Availability}

All data generated or analysed during this study are included in this published article.



\begin{thebibliography}{99}

\bibitem{raussendorf2001one}
R. Raussendorf and H. J. Briegel, Phys. Rev. Lett. \textbf{86}, 5188 (2001).

\bibitem{briegel2009measurement}
H. J. Briegel, D. E. Browne, W. D\"ur, R. Raussendorf, and M. Van den Nest, Nat. Phys. \textbf{5}, 19 (2009).

\bibitem{nielsen2006cluster}
M. A. Nielsen, Rep. Math. Phys. \textbf{57}, 147 (2006).

\bibitem{kok2007linear}
P. Kok, W. J. Munro, K. Nemoto, T. C. Ralph, J. P. Dowling, and G. J. Milburn, Rev. Mod. Phys. \textbf{79}, 135 (2007).

\bibitem{obrien2009photonic}
J. L. O'Brien, A. Furusawa, and J. Vučković, Nat. Photonics \textbf{3}, 687 (2009).

\bibitem{menicucci2006universal}
N. C. Menicucci, P. van Loock, M. Gu, C. Weedbrook, T. C. Ralph, and M. A. Nielsen, Phys. Rev. Lett. \textbf{97}, 110501 (2006).

\bibitem{gu2009quantum}
M. Gu, C. Weedbrook, N. C. Menicucci, T. C. Ralph, and P. van Loock, Phys. Rev. A \textbf{79}, 062318 (2009).

\bibitem{zhang2006continuous}
J. Zhang and S. L. Braunstein, Phys. Rev. A \textbf{73}, 032318 (2006).

\bibitem{larsen2021fault}
M. V. Larsen, C. Chamberland, K. Noh, J. S. Neergaard-Nielsen, and U. L. Andersen, PRX Quantum \textbf{2}, 030325 (2021).

\bibitem{du2025complete}
P. Du, J. Zhang, T. Zhang, R. Yang, and J. Gao, Sci. Rep. \textbf{15}, 18199 (2025).
DOI: 10.1038/s41598-025-02899-8.

\bibitem{asavanant2019generation}
W. Asavanant, Y. Shiozawa, S. Yokoyama, B. Charoensombutamon, H. Emura, R. N. Alexander, S. Takeda, J.-i. Yoshikawa, N. C. Menicucci, H. Yonezawa, and A. Furusawa, Science \textbf{366}, 373 (2019).

\bibitem{yoshikawa2016generation}
J.-i. Yoshikawa, S. Yokoyama, T. Kaji, C. Sornphiphatphong, Y. Shiozawa, K. Makino, and A. Furusawa, APL Photonics \textbf{1}, 060801 (2016).

\bibitem{larsen2019deterministic}
M. V. Larsen, X. Guo, C. R. Breum, J. S. Neergaard-Nielsen, and U. L. Andersen, Science \textbf{366}, 369 (2019).

\bibitem{menicucci2011temporal}
N. C. Menicucci, Phys. Rev. A \textbf{83}, 062314 (2011).

\bibitem{alexander2014one}
R. N. Alexander, P. Wang, N. Sridhar, M. Chen, O. Pfister, and N. C. Menicucci, Phys. Rev. A \textbf{89}, 032330 (2014).

\bibitem{ahadi2026ccr}
A. Ahadi and S. Sarshar, ``Introducing the Correlation Concentration Ratio (CCR): Quantitative Framework for Comparing Quantum Cluster States,'' arXiv:2604.23258 [quant-ph] (2026).
DOI: 10.48550/arXiv.2604.23258.

\bibitem{vidal2002computable}
G. Vidal and R. F. Werner, Phys. Rev. A \textbf{65}, 032314 (2002).

\bibitem{plenio2005logarithmic}
M. B. Plenio and S. Virmani, Quantum Inf. Comput. \textbf{7}, 1 (2007).

\bibitem{wehrl1978entropy}
A. Wehrl, Rev. Mod. Phys. \textbf{50}, 221 (1978).

\bibitem{lvovsky2009homodyne}
A. I. Lvovsky and M. G. Raymer, Rev. Mod. Phys. \textbf{81}, 299 (2009).


\bibitem{hein2006entanglement}
M. Hein, W. D\"ur, J. Eisert, R. Raussendorf, M. Van den Nest, and H. J. Briegel, in \textit{Proceedings of the International School of Physics "Enrico Fermi"} (IOS Press, Amsterdam, 2006), Vol. 162, pp. 115-218.

\bibitem{weedbrook2012gaussian}
C. Weedbrook, S. Pirandola, R. Garc\'ia-Patr\'on, N. J. Cerf, T. C. Ralph, J. H. Shapiro, and S. Lloyd, Rev. Mod. Phys. \textbf{84}, 621 (2012).

\bibitem{braunstein2005quantum}
S. L. Braunstein and P. van Loock, Rev. Mod. Phys. \textbf{77}, 513 (2005).

\bibitem{eisert2003introduction}
J. Eisert and M. B. Plenio, Int. J. Quantum Inf. \textbf{1}, 479 (2003).

\bibitem{vahlbruch2016ultra}
H. Vahlbruch, M. Mehmet, K. Danzmann, and R. Schnabel, Phys. Rev. Lett. \textbf{117}, 110801 (2016).

\bibitem{slapnicar2013symmetric}
I. Slapni\v{c}ar,``Symmetric Matrix Eigenvalue Techniques,'' in \textit{Handbook of Linear Algebra}, 2nd ed., edited by L. Hogben (Chapman and Hall/CRC, 2013).

\bibitem{west2001introduction}
D. B. West, \textit{Introduction to Graph Theory}, 2nd ed. (Prentice Hall, 2001).

\bibitem{diestel2017graph}
R. Diestel, \textit{Graph Theory}, 5th ed. (Springer, 2017).


\bibitem{vuillot2023fault}
C. Vuillot, H. Asasi, Y. Wang, L. P. Pryadko, and B. M. Terhal, Phys. Rev. A \textbf{99}, 032344 (2019).
DOI: 10.1103/PhysRevA.99.032344.

\bibitem{efron1994introduction}
B. Efron and R. J. Tibshirani, \textit{An Introduction to the Bootstrap} (Chapman \& Hall, 1994).

\bibitem{bondy2008}
J. A. Bondy and U. S. R. Murty,
\textit{Graph Theory},
Springer, 2008.

\bibitem{Baragiola2019}
B. Q. Baragiola, G. Pantaleoni, R. N. Alexander, A. Karanjai, and N. C. Menicucci, Phys. Rev. Lett. \textbf{123}, 200502 (2019).

\bibitem{gottesman2001encoding}
D. Gottesman, A. Kitaev, and J. Preskill, Phys. Rev. A \textbf{64}, 012310 (2001).

\bibitem{walshe2019continuous}
B. W. Walshe, L. J. Mensen, B. Q. Baragiola, and N. C. Menicucci, Phys. Rev. A \textbf{100}, 010301(R) (2019).

\bibitem{alexander2016scalable}
R. N. Alexander, P. Wang, N. Sridhar, M. Chen, O. Pfister, and N. C. Menicucci, Phys. Rev. A \textbf{94}, 032327 (2016).

\end{thebibliography}
\end{document}